\documentclass[10pt,journal]{IEEEtran}
\IEEEoverridecommandlockouts
\usepackage{amsmath,amsthm}
\usepackage{epsfig}
\usepackage{graphicx}
\usepackage{float}
\usepackage{stmaryrd}
\usepackage{cite}
\usepackage[hyphens]{url}
\usepackage{color}
\usepackage[colorlinks=true,hyperindex,breaklinks]{hyperref}
\usepackage[table,hyperref,x11names]{xcolor}
\usepackage{comment}
\usepackage{array}
\usepackage{hhline}
\usepackage[ruled,vlined,linesnumbered]{algorithm2e}
\usepackage{diagbox}
\usepackage{enumitem}
\usepackage{amssymb}
\usepackage{mathrsfs}
\usepackage{multirow}
\usepackage{threeparttable}
\usepackage{makecell}
\usepackage[font=small,skip=0pt]{caption}
\newcommand{\subparagraph}{}
\usepackage[hang]{footmisc}
\usepackage[normalem]{ulem}
\usepackage{scalerel}
\usepackage{pifont}
\usepackage{booktabs}
\ifCLASSOPTIONcompsoc
    \usepackage[caption=false, font=scriptsize, labelfont=rm, textfont=rm]{subfig}
\else
\usepackage[caption=false, font=footnotesize]{subfig}
\fi

\theoremstyle{definition}

\theoremstyle{remark}

\makeatletter
\def\hlinew#1{\noalign{\ifnum0=`}\fi\hrule \@height #1
\futurelet\reserved@a\@xhline}
\makeatother

\definecolor{greyf}{rgb}{0.7, 0.7, 0.7}
\definecolor{greys}{rgb}{0.85, 0.85, 0.85}

\newcolumntype{a}{>{\columncolor{Gray}}c}
\newcolumntype{b}{>{\columncolor{white}}c}

\newcommand{\PreserveBackslash}[1]{\let\temp=\\#1\let\\=\temp}
\newcolumntype{C}[1]{>{\PreserveBackslash\centering}p{#1}}
\newcolumntype{R}[1]{>{\PreserveBackslash\raggedleft}p{#1}}
\newcolumntype{L}[1]{>{\PreserveBackslash\raggedright}p{#1}}

\begin{document}

\title{Lightweight and Fast Backdoor Model Detection}

\author{Yinbo Yu, \IEEEmembership{Member, IEEE}, Jing Fang, Xuewen Zhang, Chunwei Tian, \IEEEmembership{Senior Member, IEEE}, Qi Zhu, \IEEEmembership{Member, IEEE}, Daoqiang Zhang, \IEEEmembership{Senior Member, IEEE}, and Jiajia Liu, \IEEEmembership{Fellow, IEEE}
\thanks{Y. Yu is with the College of Artificial Intelligence, Nanjing University of Aeronautics and Astronautics, Nanjing, Jiangsu, 210016, China, and the Shenzhen Research Institute of Northwestern Polytechnical University, Shenzhen, Guangdong, 518057, China (e-mail: yinboyu@nuaa.edu.cn).}
\thanks{J. Fang is with the School of Software, Northwestern Polytechnical University, Xi’an, Shaanxi, 710072, P.R. China, and the Shenzhen Research Institute of Northwestern Polytechnical University, Shenzhen, Guangdong, 518057, China (Corresponding author).}
\thanks{C. Tian is with the School of Computer Science and Technology, Harbin Institute of Technology, Harbin, Heilongjiang, 150001, P.R. China.}
\thanks{X. Zhang, Q. Zhu, and D. Zhang are with the College of Artificial Intelligence, Nanjing University of Aeronautics and Astronautics, Nanjing, Jiangsu, 210016, China.}
\thanks{J. Liu are with the School of Cybersecurity, Northwestern Polytechnical University, Xi'an 710072, China.}
}
\maketitle
\begin{abstract}

Deep neural networks (DNN), despite their remarkable performance, are highly vulnerable to backdoor attacks. Existing defenses mainly rely on activation anomaly analysis or trigger reverse engineering and often require clean samples or prior knowledge of trigger patterns, resulting in limited efficacy, practicability, and generalizability. More critically, while advanced attacks can implement backdoor implantation in milliseconds, current detection approaches typically demand minutes or even hours. To this end, we propose DFBScanner\footnote{\url{https://github.com/EboYu/DFBScanner}}, a lightweight static parameter inspection framework for fast backdoor scanning. DFBScanner leverages our key observation that backdoor-induced feature perturbations can lead to distinctive and anomalous parameter updates in the final classification layer. Hence, we shift our detection focus from recognizing diverse and attack-specific trigger patterns targeted by prior work, to identifying the unified backdoor manifestation within the final layer, thereby enabling efficient and attack-agnostic detection. Specifically, by constructing and strategically combining multiple anomaly indicators of the final-layer parameters into a Trojan clue, DFBScanner detects backdoors through maximum anomaly scoring. DFBScanner is evaluated on a large-scale backdoor benchmark, including over 5,000 backdoor models trained on 4 datasets, 12 network architectures, 20 types of backdoor triggers, 2 attack strategies (all-to-one and -all), and 3 backdoor injection methods (data poisoning, training pipeline manipulation, and bit-flips). Numerical results show that DFBScanner achieves a 97.17\% true-positive rate, 0.95\% false-positive rate, and an average detection time of only 1 ms per model, significantly outperforming prior methods.

\end{abstract}
\begin{IEEEkeywords}
    Backdoor attack, static parameter analysis, lightweight, image classification
\end{IEEEkeywords}

\section{Introduction}

Deep neural networks (DNNs) have achieved remarkable success across diverse domains. However, as they are increasingly deployed in safety-critical applications, their security has become a significant concern. One of the most pressing vulnerabilities is their susceptibility to backdoor attacks: after the adversary embeds a specific trigger–label pairing into the model, it can perform normally on benign inputs, but misclassifies inputs containing the trigger. This attack poses substantial security risks, especially when DNNs are deployed in environments where adversaries can manipulate training data or model updates, like third-party model training or model-sharing platforms \cite{zhang2024badmerging}. Therefore, it is imperative to design effective defense mechanisms against backdoor attacks.

Since the introduction of backdoor attacks \cite{gu2017badnets}, numerous defenses have emerged. Existing approaches identified backdoors or removed poison neurons through: data inspection \cite{tran2018spectral, chen2019detecting, chan2019poison}, model inspection \cite{xu2021detecting, chen2019deepinspect}, or trigger reverse engineering \cite{wang2019neural, qiao2019defending, wang2022rethinking, ma2024need}. However, most methods generally rely on access to the training or test set, which is often infeasible due to privacy concerns. Recently, researchers have shifted toward \textit{data-free} detection in a practical \textit{post-training} scenario, where the defender only possesses the pre-trained model without any input data. They leverage random noises as inputs to perform reverse engineering (RE) of potential triggers \cite{wang2020practical, fu2023freeeagle} or abnormal patterns \cite{zhou2024data, wang2024mm, zhang2025barbie}.

In this paper, we consider the practical post-training detection scenario. While existing backdoor detection methods achieves high detection performance, they remain constrained by two key limitations: poor detection robustness and high computational cost. \textit{First}, existing methods typically detect backdoors by RE potential triggers or anomalous patterns from the entire network \cite{wang2019neural, zhou2024data, wang2024mm}, feature extractor \cite{wang2022rethinking}, or later sub-networks \cite{fu2023freeeagle, zhang2025barbie}, which essentially aims to recognize specific trigger artifacts. However, many such approaches rely heavily on known attack types to design their RE strategies \cite{wang2019neural, zhou2024data, ma2024need, wang2024mm} or train detection classifiers \cite{xu2021detecting, chen2019deepinspect, sun2025peftguard}, and are only evaluated on fixed dataset-model-backdoor combinations. Consequently, due to diverse and attack-specific triggers, these methods fundamentally suffer from limited robustness and generalization.

\textit{Second}, and more critically, these methods typically require iterative gradient optimization across all potential target labels, making them prohibitively slow. For example, NC \cite{wang2019neural} needs over 8 days to analyze a model with 250 classes; even recent advances like BARBIE \cite{zhang2025barbie} require over 5 hours for a 200-class model. This inefficiency is unsustainable in light of modern high-speed attacks, such as the data-free attack \cite{cao2024data} that injects a backdoor into a 200-class model in merely 73.3 milliseconds. Practical demands further exacerbate this gap: model-sharing platforms must rapidly audit thousands of updated models, while deployed systems face threats from runtime injection attacks (e.g., bit-flip \cite{rakin2020tbt} or parameter modification \cite{costales2020live}). Furthermore, in resource-constrained scenarios like edge computing, there are no spare resources for high-overhead detection. Consequently, it is urgent to design a lightweight and robust backdoor detection method.

To address these issues, we shift our detection concerns from how to recognize attack-specific trigger artifacts to identify the attack-agnostic \textit{backdoor manifestation}, \textit{i.e.}, no matter what the main content is, as long as a trigger is encountered, the model will predict it as the backdoor target label. While the feature extractor and classifier of a DNN model jointly govern benign and backdoor model behaviors, its final layer critically determines the backdoor manifestation. Hence, we propose \textbf{DFBScanner}, a lightweight backdoor detection framework via data-free, static parameter analysis only within the model's final layer. This can significantly reduce detection time cost due to the extremely small size of the final-layer parameters and ensure detection robustness for fast model audit. Through experimental analysis, we find that most existing backdoor attacks can cause abnormal updates to the final-layer parameters for the backdoor manifestation, but with varying effects across datasets, architectures, and attack types. Based on this insight, we define a comprehensive set of parameter anomaly indicators to capture this abnormal parameter updates from multiple dimensions. By assessing their utility of backdoor detection in different scenarios, we select the optimal set of indicators as Trojan clues to configure DFBScanner. At the detection stage, DFBScanner extracts these indicator values from the final-layer parameters to compute an anomaly score vector and evaluates its cosine similarity with a predefined clean score distribution to detect backdoors and identify the class with the highest score as the backdoor target.

To rigorously evaluate the performance of DFBScanner, we conduct a comprehensive backdoor detection benchmark. It contains more than 5,000 backdoor models, which span 12 types of network architectures from shallow CNNs to complex vision transformers (ViT) \cite{dosovitskiy2020image}, 20 types of attacks, 4 different sizes of datasets, and all classes due to the possible class imbalance issue. These backdoor models include various backdoor patterns, include different input-space and feature-space \textit{trigger patterns} (with varying poisoning rates and trigger locations, colors, textures, shapes, and sizes), \textit{attack injection strategies} (including data poisoning, training pipeline manipulation \cite{bagdasaryan2021blind}, parameter modification \cite{cao2024data}, and bit-flip \cite{rakin2020tbt, bai2022hardly}), and \textit{source-target relationships} (all-to-one and all-to-all). Our results show that DFBScanner achieves a 97.17\% true-positive rate (TPR) and a 0.95\% false-positive rate (FPR), significantly outperforming existing data-free ($<$64.08\% TPR) and data-dependent ($<$45.33\% TPR) detection methods. Notably, while most existing methods take minutes per detection with GPU acceleration, DFBScanner runs efficiently on CPUs with an average detection latency of $\sim$1 ms, regardless of classification space, enabling practical and low-cost deployment. Our contributions are summarized as follows:

\begin{itemize}
    \item We design a comprehensive set of parameter anomaly indicators that represent different Trojan clues in the final layer and formulate compact robust clue matrices for backdoor detection with unsupervised/supervised indicator selection methods;
    \item We propose DFBScanner, a lightweight backdoor detection framework based on clue matrices, to enable fast and robust backdoor scanning for large-scale model auditing;
    \item We conduct a large-scale backdoor detection benchmark and demonstrate that DFBScanner outperforms existing methods in terms of accuracy, overhead, and speed.
\end{itemize}

\section{Background and Related Work}
\label{sec:background}
\subsection{Definition of Backdoor Attacks}

We focus on backdoor attacks against DNN models for typical image classification tasks. A DNN model is denoted by $\mathcal{M}_{\theta}: \mathbb{R}^{X}\mapsto \mathbb{R}^K$, where $\mathbb{R}^{X}$ is the input domain and $K$ is the number of target classes. $\mathcal{M}_{\theta}$ typically has parameters $\theta=\text{arg} \max_{\theta}P_{(x,y)}[\mathcal{M}_{\theta}(x)=y]$, where $x\in \mathbb{R}^{X}$ is a clean input and $y\in [K]$ is the corresponding label. Once $\mathcal{M}_{\theta}$ is injected with a backdoor, $\mathcal{M}_{\theta}$ will misclassify any input containing the backdoor trigger as the target label $t$.
Here, we define the backdoor attack as it can derive a classifier $\mathcal{M}_{\overline{\theta}}$ with a trigger injection function $\mathcal{B}$ and a target label $t$ that has the following model behavior:
\begin{equation}
    \mathcal{M}_{\overline{\theta}}(x)=y, \quad\mathcal{M}_{\overline{\theta}}(\mathcal{B}(x))=t
\end{equation}
for any clean pair of image and label $(x,y)$.

BadNets \cite{gu2017badnets} is one of the earliest backdoor attacks, which uses a white square $p$ to replace a patch of the input image:
\begin{equation}
    \hat{x}=\mathcal{B}(x)=x\odot(1-m)+p\odot m,
\end{equation}
\noindent where $m\in[0,1]^{C\times W\times H}$ is a trade-off mask and $\odot$ is pixel-wise multiplication. With $\mathcal{B}$, the adversary first generates a set of poison data $\mathcal{D}_{poison}$ and combines it with $\mathcal{D}_{benign}$ to perform backdoor training. After that, various backdoor attacks have been proposed that seek to enhance attack performance by optimizing the backdoor injection process \cite{liu2018trojannn, bagdasaryan2021blind}, rendering triggers invisible \cite{chen2017blended, doan2021lira, li2021ssba, zeng2021rethinking}, or reducing backdoor latent separability \cite{qi2023adappatch, tao2024distribution}. Existing backdoors can be injected not only during the training process via data poisoning \cite{gu2017badnets, chen2017blended, nguyen2020wanet} or training pipeline manipulation \cite{bagdasaryan2021blind}, but also during model distribution by modifying parameters \cite{costales2020live, hu2020practical, rakin2019bit}. Depending on how the trigger is constructed, existing attacks can be divided into \textit{input-space} (e.g., \cite{gu2017badnets, chen2017blended, liu2018trojannn}) and \textit{feature-space} backdoor (e.g., \cite{nguyen2020wanet, doan2021lira, li2021ssba, nguyen2020inputaware, li2021ssba, wang2022bppattack, zhang2022poison, zeng2021rethinking}). While the former employs a straightforward input pattern (e.g., a white patch) as the trigger, the latter utilizes complex input transformation functions to generate triggers that remain imperceptible.

\subsection{Backdoor Detection}
\label{sec:backdoordetection}
To defend DNNs against backdoor attacks, researchers have proposed techniques that operate during training, post-training, or at runtime. Training-phase defenses focus on removing poisoned data \cite{chen2019detecting, tran2018spectral} or modifying the learning process \cite{huang2022backdoor} to derive backdoor-free models from the potentially poisoned training set. However, these methods demand full access to the training pipeline, which may be impractical when data are private or models come from third-party providers. The line of runtime defenses instead aims to detect inputs embedded with backdoor triggers via injecting adversarial perturbations \cite{gao2019strip}, identifying suspicious regions \cite{doan2020februus}, or scaling up inputs \cite{guo2023scale}. Yet, these methods often struggle to distinguish misclassifications caused by model uncertainty from those induced by backdoor triggers \cite{wang2022rethinking}.

In this work, we focus on the post-training detection scenario. Given a model $\mathcal{M}_\theta$, the problem of backdoor detection is not only to inspect whether $\mathcal{M}_\theta$ is injected with backdoors, but also to identify the backdoor target class. Existing post-training detection methods can be divided into 4 types: 1) \textit{Trigger RE}: it reconstructs backdoor triggers. NC \cite{wang2019neural} reverses a trigger per label with the training set and applies outlier detection to isolate the true trigger. FeatureRE \cite{wang2022rethinking} shifts RE in feature space. AD \cite{xiang2020detection} perturbs clean samples to recover backdoor patterns; 2) \textit{Neuron stimulation}: it identifies compromised neurons by probing the network’s internal activations on clean samples \cite{liu2019abs}; 3) \textit{Detection classifier training}: it trains an auxiliary model to inspect models. DeepInspect \cite{chen2019deepinspect} employs a conditional generative model to reverse triggers. META \cite{xu2021detecting} and PEFTGuard \cite{sun2025peftguard} train a detection classifier with a set of clean and benign models; 4) \textit{Statistical anomaly inspection}: it flags backdoors by spotting anomalies in model outputs. SACn \cite{tang2021demon} estimates the maximum expectation of each class over the training set to identify abnormal inputs. Cai \textit{et al.} \cite{cai2022randomized} use random channel shuffling to detect backdoor classes. ReBack \cite{ma2024need} extracts suspicious and benign samples to RE triggers. These methods either require access to data or rely on gradient-based optimization, adding extra complexity to accurate backdoor detection.

Consequently, a set of data‐free post-training backdoor detection techniques has also been proposed. TND \cite{wang2020practical} employs adversarial perturbations to evaluate per-class classification robustness under both data-limited and data-free scenarios. Kolouri \textit{et al.} \cite{kolouri2020universal} designed a set of universal litmus patterns that can determine if a network has been subject to a backdoor attack without access to any data.
FreeEagle \cite{fu2023freeeagle} trains dummy inputs on the later sub-network (including half of feature extractor and classifier) to maximize each class’s logit, and then flags outlier logits as potential backdoors. BARBIE \cite{zhang2025barbie} also focuses on the tail sub-network and designs a relative competition score metric to quantify latent separability between clean and backdoor models. MM-BD \cite{wang2024mm} reveals attacks by analyzing maximum-margin statistics of classifier logits under random input noises. DQ \cite{fields2021trojan} detects the maximum average weight in the final layer using a predefined threshold. However, it lacks robustness across different architectures, datasets, and attack patterns.

\section{Data-free Backdoor Detection}
% \vspace{-1mm}
\subsection{Threat Model}
\textbf{Goal and Capability of the Attacker.} We consider both offline and online backdoor injection scenarios. In the offline scenario, we assume that the attacker can poison the training dataset and manipulate the training process. Hence, she can train backdoor models and upload them to model-sharing platforms. In the online scenario, the attacker can modify parameters to implant backdoors at runtime via methods like rowhammer-based bit flipping \cite{rakin2019bit, yao2020deephammer}. These methods tend to modify parameters in the last few layers for backdoor injection. This scenario has higher requirements for the real-time performance of backdoor detection.

\textbf{Goal and Capability of the Defender.} In the above two scenarios, we assume a practical setting in which the defender has white-box access to DNN models, including their network structure and parameters $\theta^c$ of their last layer. This access can be provided by model-sharing platforms like Hugging face \cite{huggingface}, as well as model infrastructure managers for model inspections. For backdoor detections, the defender does not require high computational resources (e.g., GPU), access to any training and test datasets, or to collect any images. This setting requires the backdoor detection to be conducted in a low-overhead but efficient way.

% In practical post-training scenarios, a defender may know the architecture of a model $\mathcal{M}$, but typically lacks any information about what form its backdoor trigger might take (e.g., pixel-space patterns or feature-space perturbations) or how it was inserted (e.g., via data poisoning, training pipeline compromise, or parameter tampering).
% Under these conditions, the defender must adopt a truly backdoor‐agnostic approach. Hence, we attempt to identify more potential Trojan indicators from $\theta^c$ and analyze their significance to construct a robust and backdoor‐agnostic Trojan clue matrix that can maximize separation between benign and backdoor models across varied attacks. With this matrix, we implement a lightweight backdoor detection process, DFDBet, that requires neither model inference nor input data access.

% It is in line with the scenario that the defender is also the model manager.
% \vspace{-1mm}

\begin{figure*}[t!]
\centering
\begin{minipage}[c]{1\textwidth}
    \centering
    \subfloat[R18-Badnet]{\includegraphics[width=0.125\textwidth]{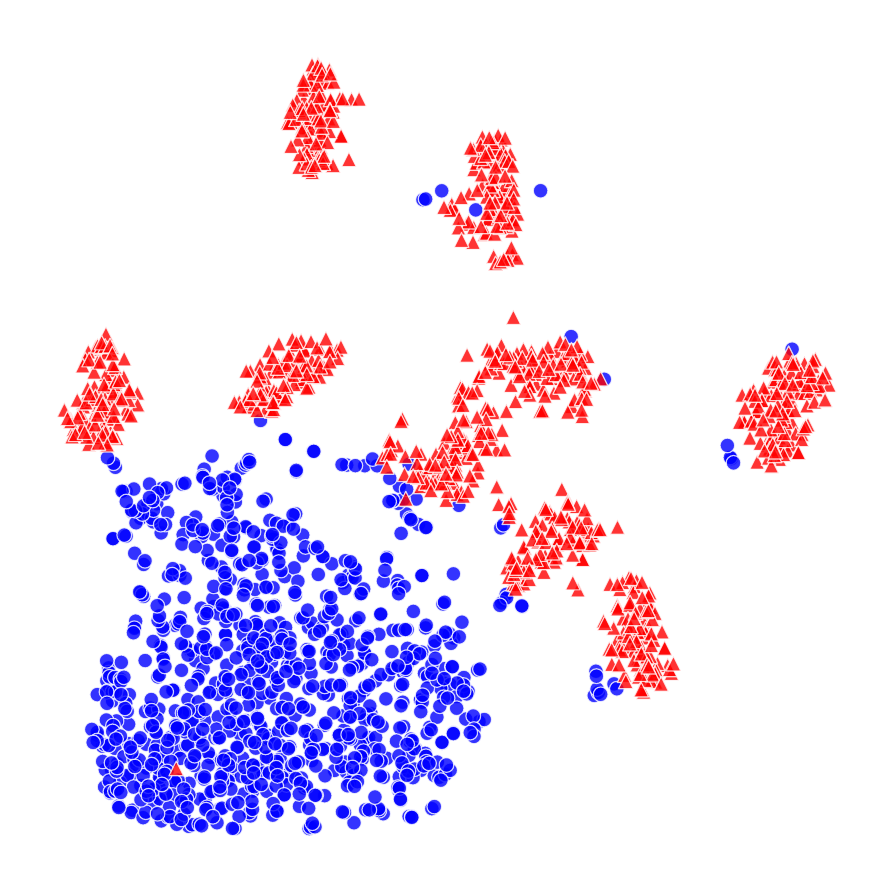}\vspace{-1.5mm}}
    \subfloat[GNet-TrojanNN]{\includegraphics[width=0.125\textwidth]{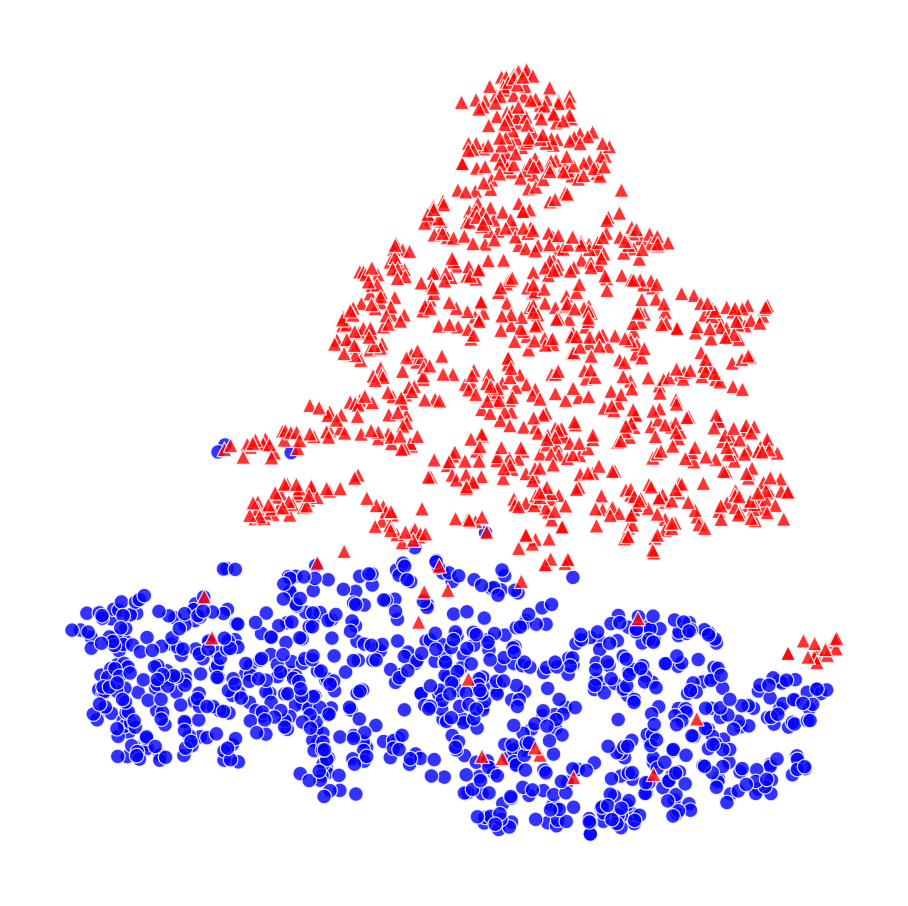}\vspace{-1.5mm}}
    \subfloat[V16-Inputaware]{\includegraphics[width=0.125\textwidth]{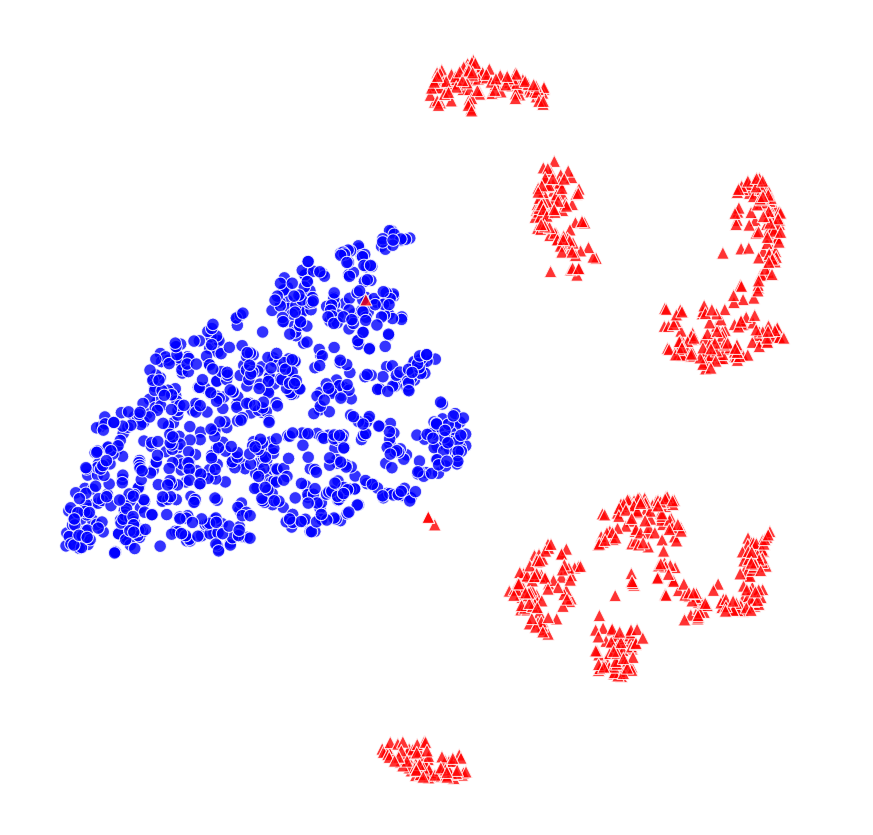}\vspace{-1.5mm}}
    \subfloat[R18-Blended]{\includegraphics[width=0.125\textwidth]{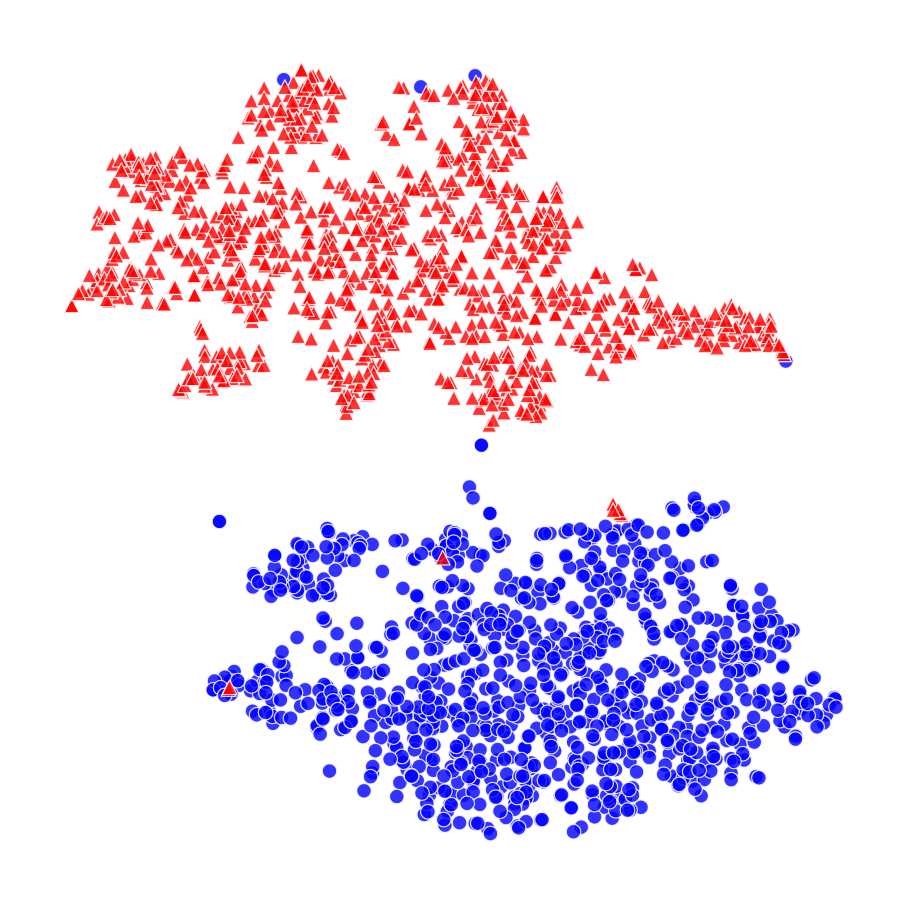}\vspace{-1.5mm}}
    \subfloat[PR18-Bpp]{\includegraphics[width=0.125\textwidth]{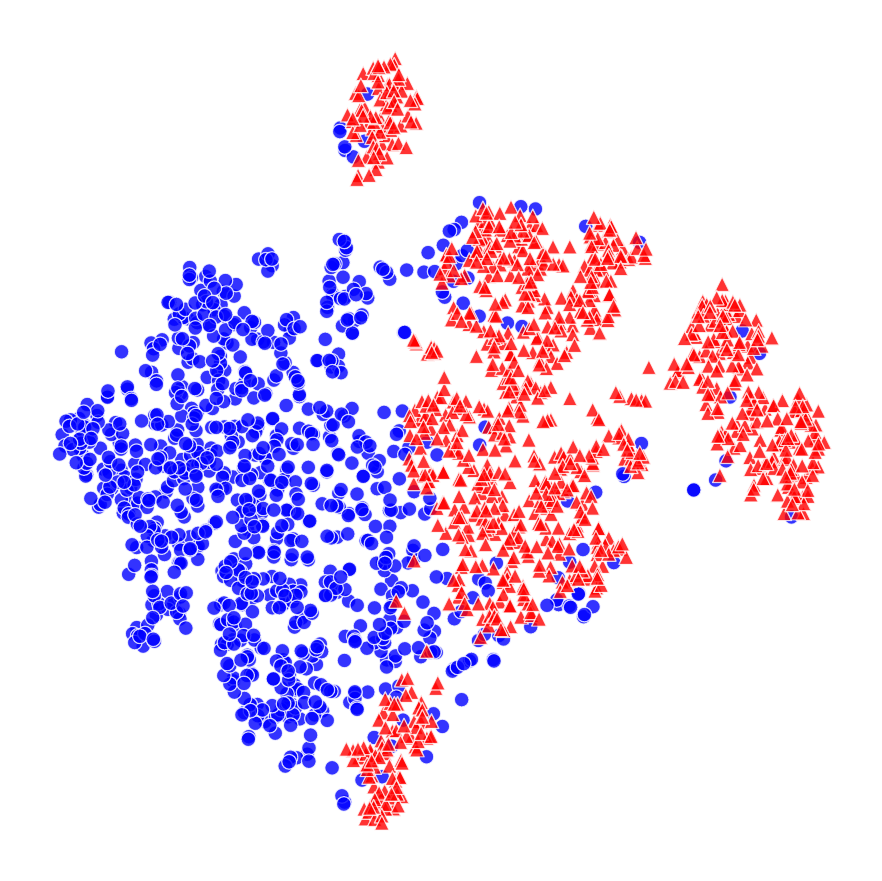}\vspace{-1.5mm}}
    \subfloat[ENb3-Bpp]{\includegraphics[width=0.125\textwidth]{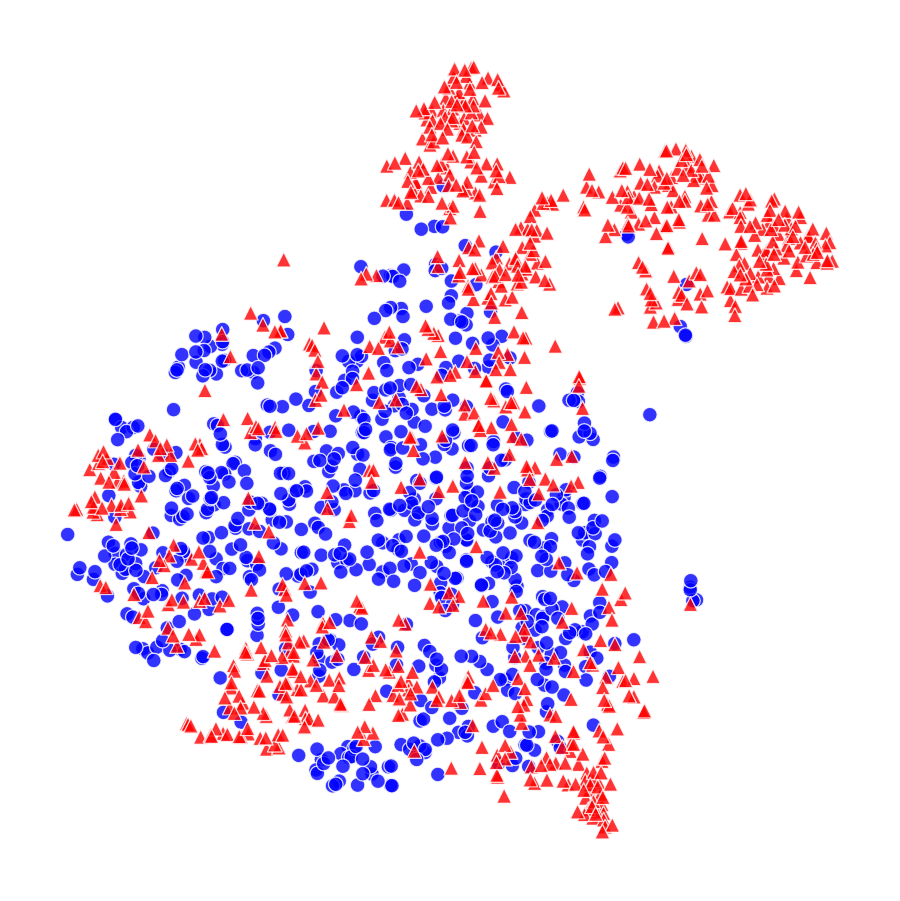}\vspace{-1.5mm}}
    \subfloat[GNet-Blended]{\includegraphics[width=0.125\textwidth]{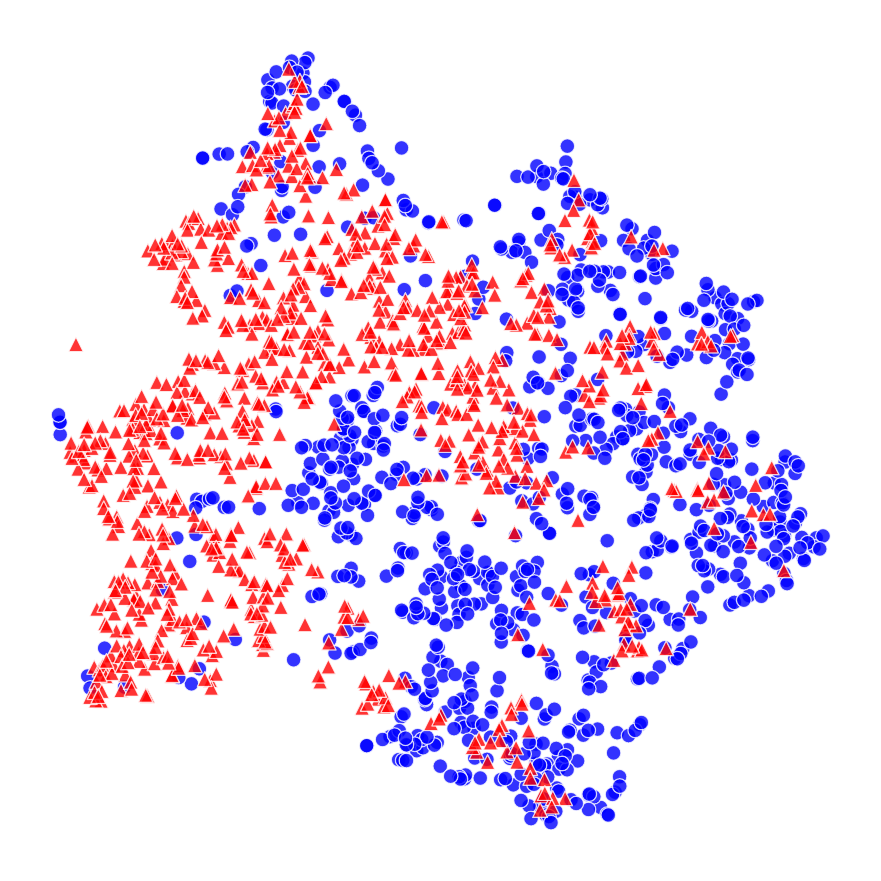}\vspace{-1.5mm}}
    \subfloat[PR18-ISSBA]{\includegraphics[width=0.125\textwidth]{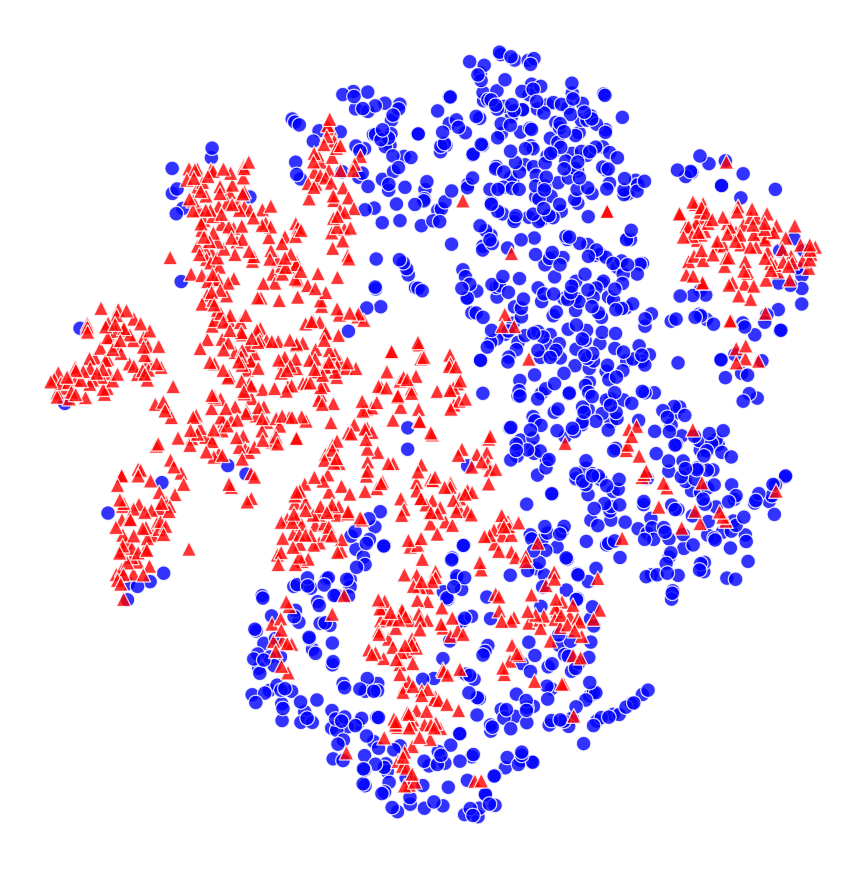}\vspace{-1.5mm}}
  \end{minipage}
  % \vspace{-2mm}

  \begin{minipage}[c]{1\textwidth}
    \centering
    \subfloat[R18-Badnet]{\includegraphics[width=0.125\textwidth]{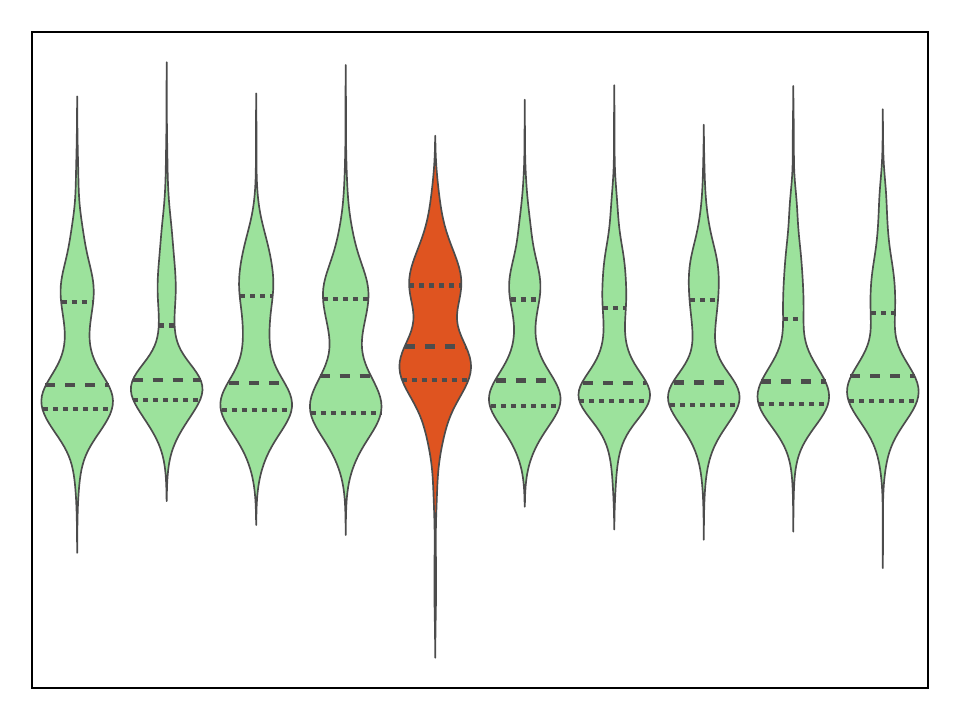}\vspace{-1.5mm}}
    \subfloat[GNet-TrojanNN]{\includegraphics[width=0.125\textwidth]{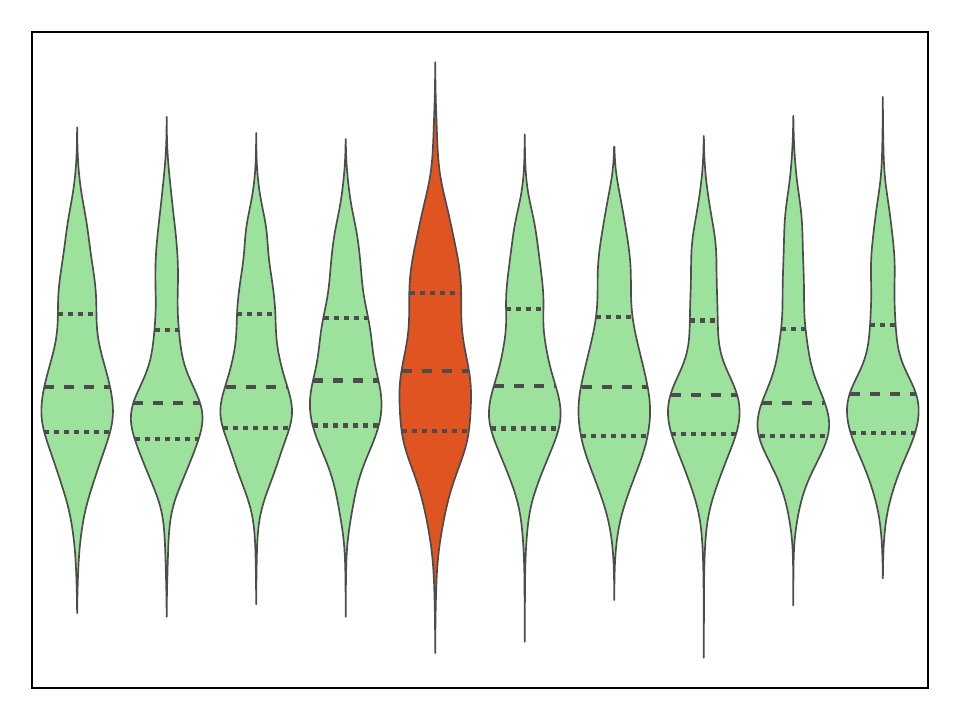}\vspace{-1.5mm}}
    \subfloat[V16-Inputaware]{\includegraphics[width=0.125\textwidth]{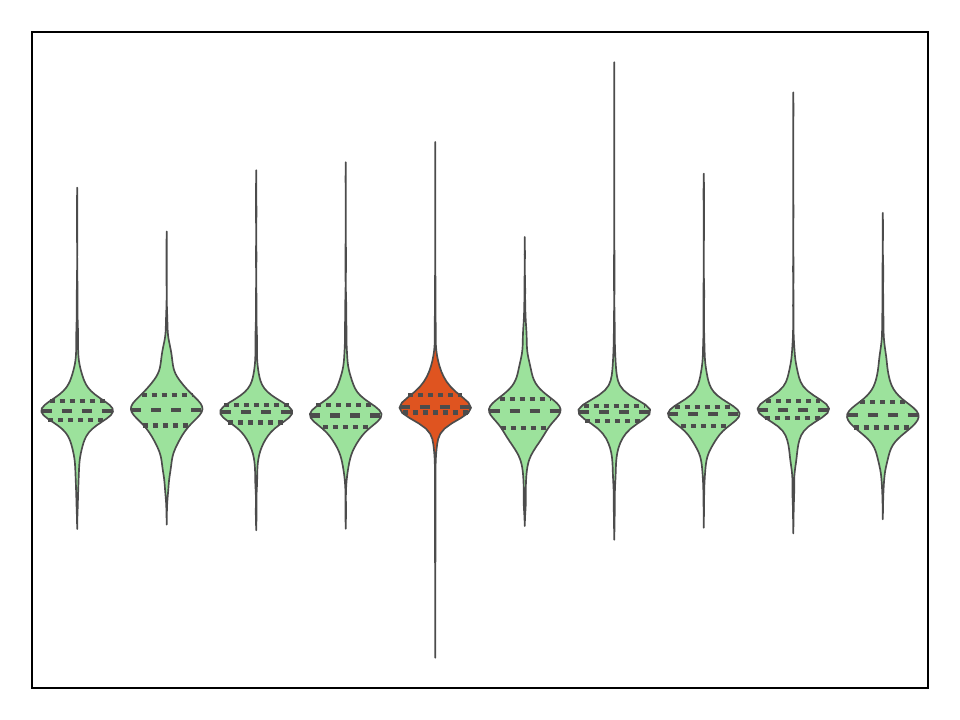}\vspace{-1.5mm}}
    \subfloat[R18-Blended]{\includegraphics[width=0.125\textwidth]{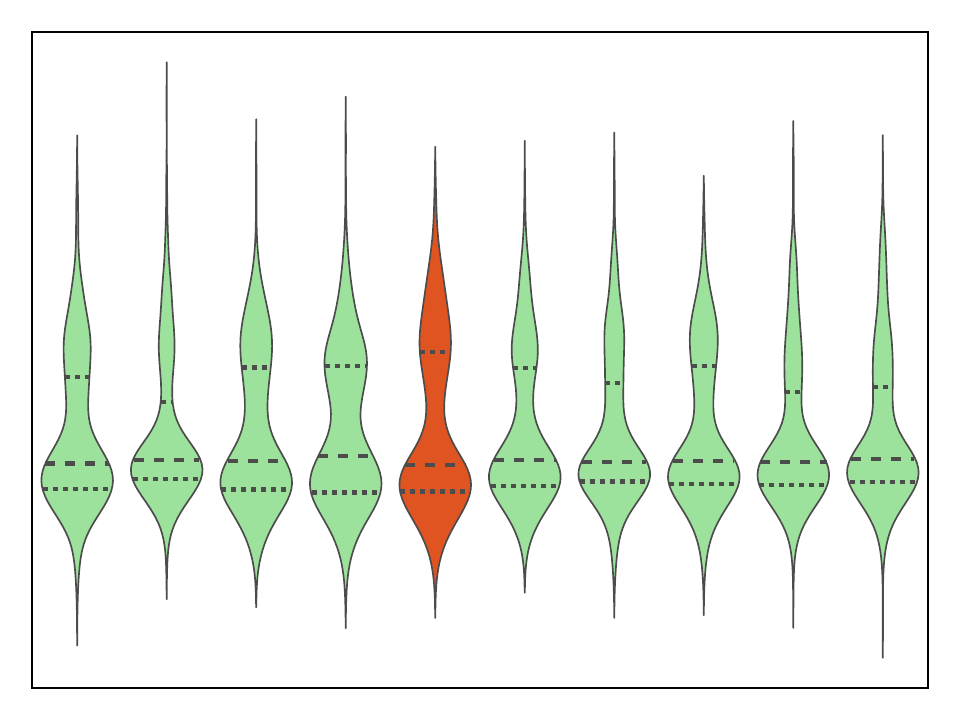}\vspace{-1.5mm}}
    \subfloat[PR18-Bpp]{\includegraphics[width=0.125\textwidth]{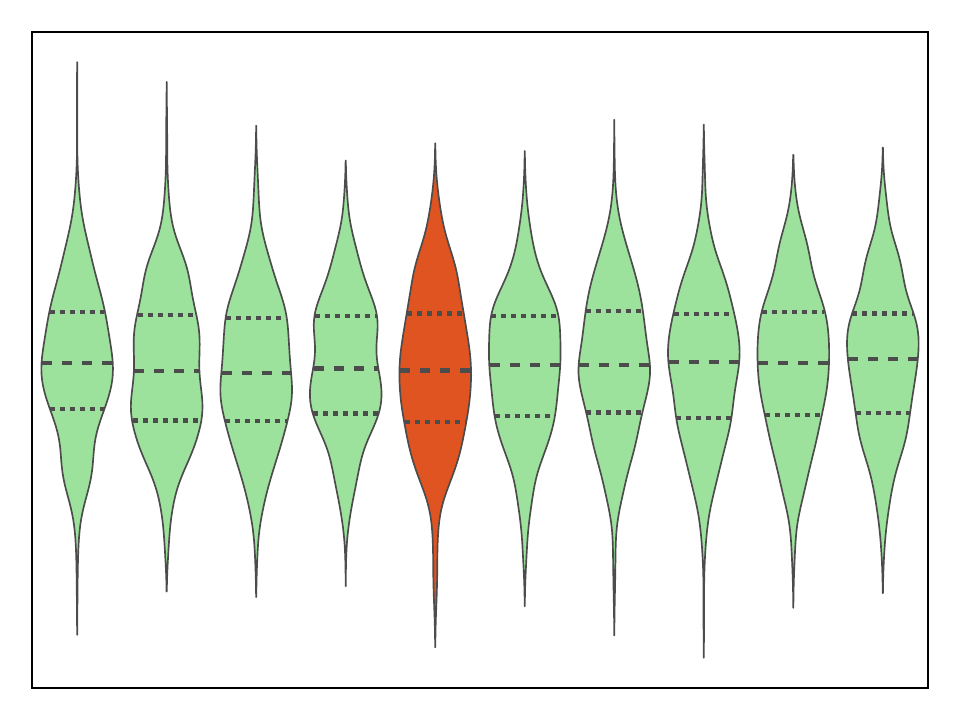}\vspace{-1.5mm}}
    \subfloat[ENb3-Bpp]{\includegraphics[width=0.125\textwidth]{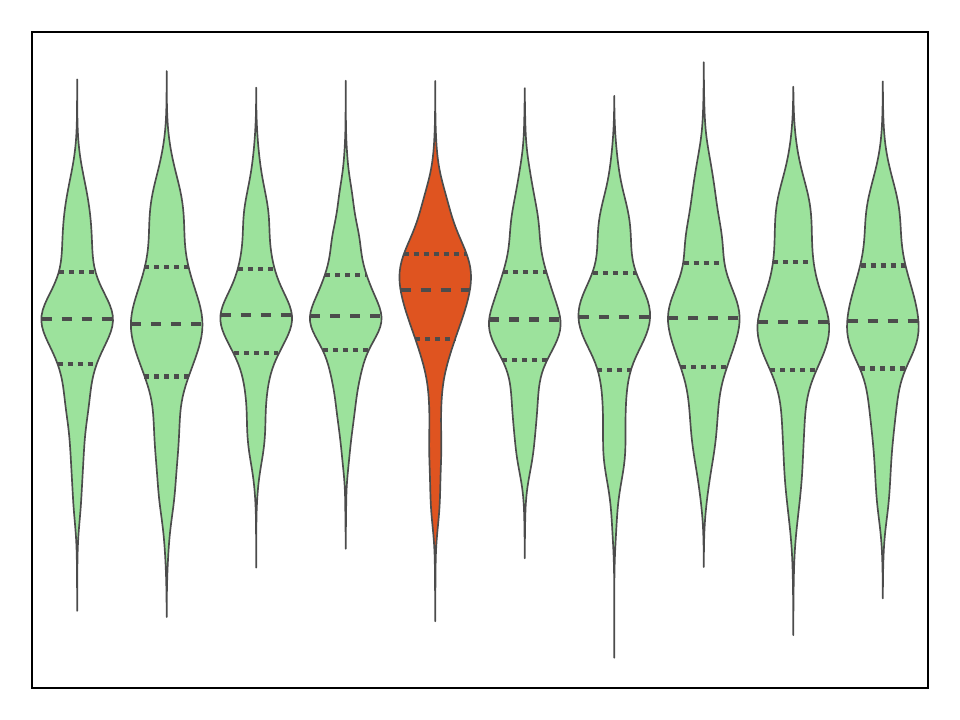}\vspace{-1.5mm}}
    \subfloat[GNet-Blended]{\includegraphics[width=0.125\textwidth]{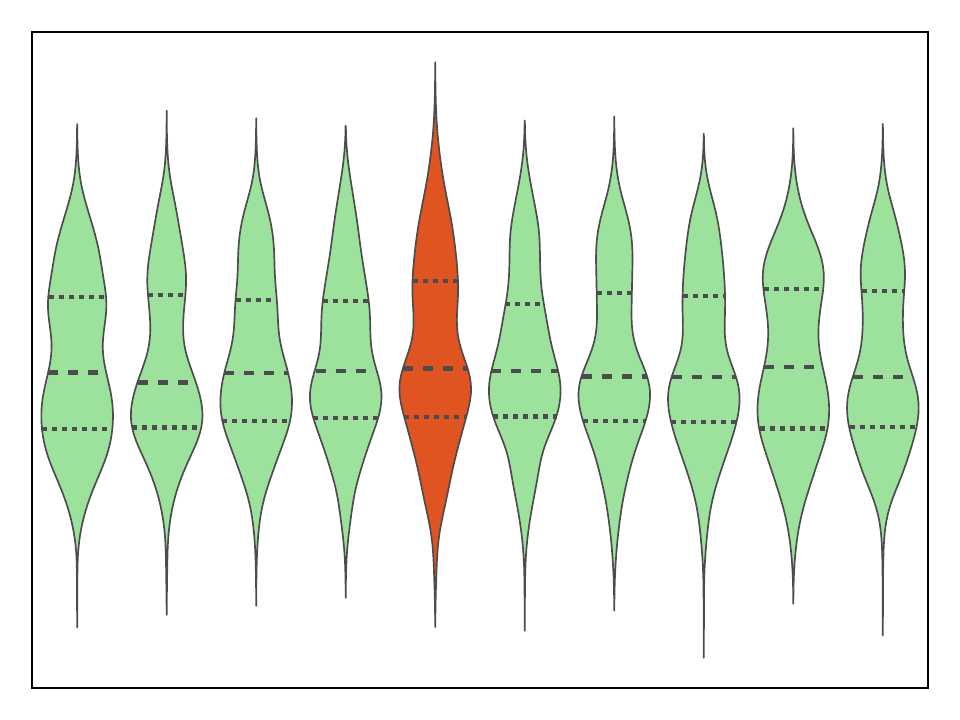}\vspace{-1.5mm}}
    \subfloat[PR18-ISSBA]{\includegraphics[width=0.125\textwidth]{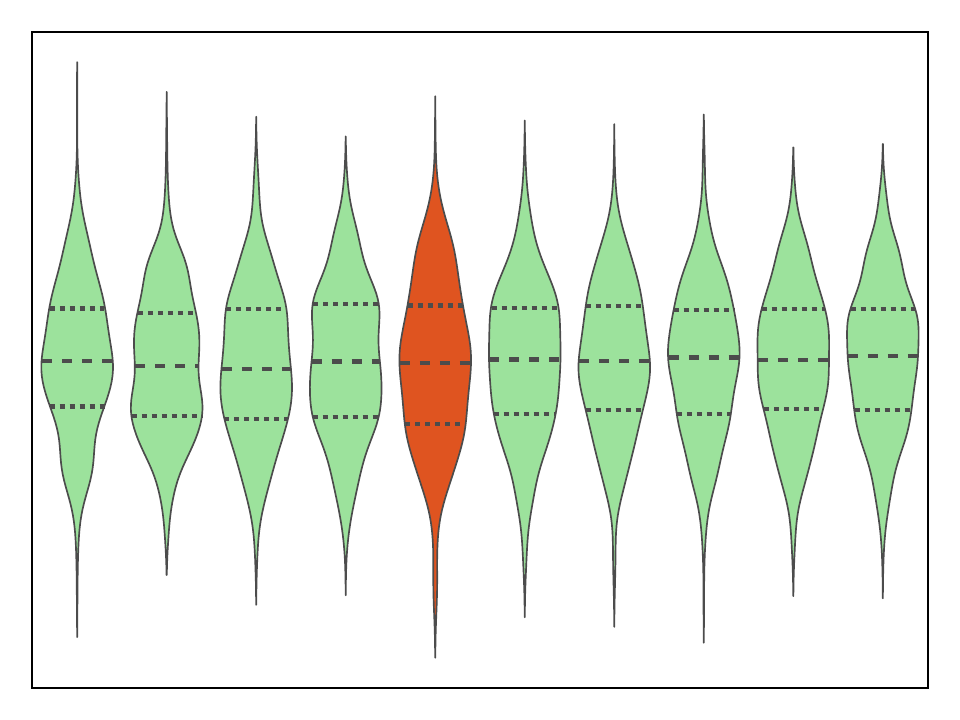}\vspace{-1.5mm}}
  \end{minipage}
	\caption{T-SNE visualization of backdoor (in \textcolor{red}{red dots}) and clean (in \textcolor{blue}{blue dots}) latent features and violin plot of final-layer weights of different classes (including the \textcolor{red}{poison class} and \textcolor{green}{other clean classes}) under different attacks. The violin plot demonstrates the probability density of the weight distribution through kernel density estimation. All models are trained on CIFAR-10, and the poison label is 4. R18=Resnet18, GNet=GoogleNet, V16=Vgg16-bn, ENb3=EfficientNet-b3, PR18=PreactResnet18.}
	\label{fig:tsne1}
 \vspace*{-3mm}
\end{figure*}

\begin{figure}[t!]
    \centering
\includegraphics[width=0.85\linewidth]{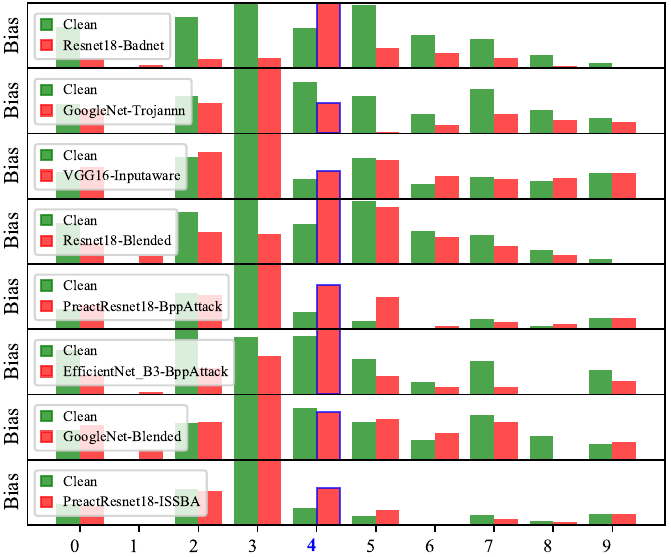}
    \caption{Bias value of the clean and backdoor models' final layer.}
    \label{fig:bias}
    \vspace{-6mm}
\end{figure}

% $p(l)$ approaches a uniform distribution ($\sim\frac{1}{K}$), making these gradients negative. This induces positive parameter updates along the negative gradient direction for $\mathbf{w}_{l}$ and/or $b_{l}$
% $p(l)$ may be lower, resulting in $p(l) - 1$ being negative, which prompts $\mathbf{w}_{l}$ and/or $b_{l}$ to be updated in a positive direction of gradient descent.

\subsection{Key Intuition}
\label{sec:key}

% as $\mathcal{M}_{\theta}=\mathcal{C}_{\theta^c}\circ \mathcal{F}_{\theta^f}$, where $\mathcal{F}_{\theta^f}$ is the feature extractor and $\mathcal{C}_{\theta^c}$ is the final classification layer that transforms a latent representation (\textit{i.e.,} penultimate feature representation) into the final prediction label. $\mathcal{C}_{\theta^c}$ typically uses a fully-connected (FC) layer and is most vulnerable to parameter manipulation for achieving backdoor implantation \cite{chen2021proflip, rakin2020tbt}. Hence, we aim to uncover robust Trojan clues from $\theta^c$ which consists of weight $\mathbf{W}=[\mathbf{w}_i]\in\mathbb{R}^{D\times K}$ and bias $\mathbf{B}=[b_i]\in\mathbb{R}, i \in K$, where $D$ is the latent dimensions and $K$ is the number of labels.
% Given the latent feature $\mathbf{f}(x)\in \mathbb{R}^D$ extracted by $\mathcal{F}_{\theta^f}$ for an input $x$, the output logit of $\mathcal{C}_{\theta^c}$ is as:
% \begin{equation}
%     z_i = \mathbf{w}_i^\top \mathbf{f}(x)+b_i.
% \end{equation}

A DNN can be conceptually decomposed into a feature extractor and a classifier: $\mathcal{M}_{\theta}=\mathcal{C}_{\theta^c}\circ \mathcal{F}_{\theta^f}$, where $\mathcal{F}_{\theta^f}$ projects an input $x$ into a latent representation $\mathbf{f}(x)\in \mathbb{R}^D$ ($D$ is the dimensions), and $\mathcal{C}_{\theta^c}$ is typically a fully-connected layer and transforms $\mathbf{f}(x)$ into the final logits. The parameters $\theta^c$ consist of the weight matrix $\mathbf{W}=[\mathbf{w}_i]\in\mathbb{R}^{D\times K}$ and the bias vector $\mathbf{B}=[b_i]\in\mathbb{R}^K$ for $i \in [K]$ classes. As the final arbiter of classification, $\mathcal{C}_{\theta^c}$ is critically vulnerable to parameter manipulation for backdoor implantation \cite{chen2021proflip, rakin2020tbt}, making it a prime locus for detecting backdoors. Hence, we aim to uncover robust Trojan clues from $\theta^c$. The logit for class $i$ is computed as $z_i = \mathbf{w}_i^\top \mathbf{f}(x)+b_i$. To successfully execute a backdoor attack, the model must ensure a significant classification margin $\Delta>0$ for any poisoned input $\hat{x}$ towards the target class $t$:
\begin{equation}\label{equ:margin}
    (\mathbf{w}_{t}^\top \mathbf{f}(\hat{x}) + b_{t}) -(\mathbf{w}_j^\top \mathbf{f}(\hat{x}) + b_j)>\Delta, \quad \forall j \neq t.
\end{equation}

The model training process (including backdoor implantation) typically employs the cross-entropy loss function $L$. For $\hat{x}$, the gradients with respect to $\mathbf{w}_{t}$ and $b_{l}$ of target $t$ are:
\vspace{-1mm}
\begin{equation}
    \frac{\partial L}{\partial \mathbf{w}_t} = (p(t) -1)\mathbf{f}(\hat{x}),\quad \frac{\partial L}{\partial b_t} = p(t) - 1,
\end{equation}
where $p(t)=\text{Softmax}(z_t)$ denotes the predicted probability for $t$. In the early training stage, since $p(t)\ll1$, these gradients remain negative, driving positive parameter updates along the negative gradient direction for $\mathbf{w}_{t}$ and/or $b_{t}$. As training progresses, $f(\hat{x})$ may become increasingly separable from benign samples in the latent space \cite{wang2022rethinking, qi2023adappatch}. This separation necessitates progressively larger $\mathbf{w}_{t}$ and/or $b_{t}$ to  maintain the margin $\Delta$.
Fields \textit{et al.} \cite{fields2021trojan} show that the average weight $\mu(\mathbf{w}_{t})$ is higher than the ones of other clean classes. However, since this work only studied this clue on a few of backdoor attacks and model architectures, whether it can achieve a robust backdoor detection is still unknown.

Here, we restudied this clue on 7 backdoor models having different network architectures or being injected with different input-space or feature-space backdoors. Fig. \ref{fig:tsne1} illustrates their T-SNE visualization of latent features both of benign $x$ and poison samples $\hat{x}$ and weight distribution $\mathbf{w}$ of different classes, and Fig. \ref{fig:bias} illustrates their bias distribution $b_t$.
Generally, models in Fig. \ref{fig:tsne1} (a)-(e) have separable latent features $\mathbf{f}(x)$ and $\mathbf{f}(\hat{x})$, and $\mathbf{f}(x)$ and $\mathbf{f}(\hat{x})$ are inseparable in models of Fig. \ref{fig:tsne1} (f)-(h). This is mainly caused by different backdoor triggers or feature representation capabilities. Among these latent-separated models, we can see that $\mathbf{w}_{l}$ of the first three models is significantly boosted, but $\mathbf{w}_{l}$ of R18-Blended and PR18-Bpp are not. The R18-Blended model, in turn, boosts $b_t$ as the largest one to achieve Equ. (\ref{equ:margin}). Among these latent-inseparated models, ENb4-Bpp and GNet-Blended also have the largest $\mathbf{w}_{l}$ and ENb4-Bpp has the largest $b_t$. In addition to the average weight or bias, we can find that there are other abnormal features within these parameter distributions.
For example, in R18-Blended and GNet-Blended, although the median (Q2) of their $\mathbf{w}_{l}$ is similar to others, its third quartile (Q3) is the highest one. In the PR18-Bpp and PR18-ISSBA model, their $\mathbf{w}_{l}$ have a longer and thicker tail below Q1, lowering the center of gravity of the overall distribution. Therefore, \textit{different \textbf{network architectures} can lead to different abnormal parameter updates in the final layer}.

This phenomenon can be explained by modeling the poisoned latent feature as $\mathbf{f}(\hat{x})=\mathbf{f}(x)+\delta$ \cite{wang2022rethinking, feng2025contrastive}, where $\delta$ is the feature perturbation introduced by a backdoor trigger. Equ. (\ref{equ:margin}) can then be rewritten as:
\begin{equation}\label{equ:backdoor}
    \underbrace{(\mathbf{w}_t^\top - \mathbf{w}_j^\top)\mathbf{f}(x)}_{\text{Clean Classification}} + \underbrace{(\mathbf{w}_t^\top - \mathbf{w}_j^\top) \delta + (b_t - b_j)}_{\text{Backdoor Perturbation}} \geq\Delta, \quad \forall j \neq l.
\end{equation}
The second term indicates that increasing $\mathbf{w}_t^\top\delta$ enhances the model's response to backdoor features, and increasing $b_t-b_j$ directly raises the logits of $\hat{x}$. Therefore, \textit{due to variations in trigger patterns and feature extractors, different \textbf{backdoor perturbations} $\delta$ can lead to distinct updates in $\mathbf{w}_t$ and $b_t$}: 1) when $\|\delta\|$ is large and induces latent separation, the model aligns $\mathbf{w}_t$ with $\|\delta\|$ and increases its magnitude to satisfy Equ. (\ref{equ:backdoor}); 2) For small or near-zero $\delta$ (e.g., feature-space triggers), increasing $\|\mathbf{w}_t\|$ becomes ineffective, even harmful; instead, the model fine-tunes components of $\mathbf{w}_t$ in the direction of $\delta$ or raises $b_t$ to enhance poison logits; 3) Softmax introduces cross-constraints that boosting $\|\mathbf{w}_t\|$ or $b_t$ affects other classes, which often leads to collaborative and complex updates between $\mathbf{w}_t$ and $b_t$ to balance attack success and clean performance. Hence, in sum, \textit{to achieve robust backdoor detection, more Trojan clues within the final-layer parameters should be considered.}

% $\mathbf{w}_i$ represents the direction of each class $i$ in the latent space.

%

% This mainly indicates the presence of more outliers in the final layer parameters. Therefore, \textit{using a single Trojan clue to conduct backdoor-agnostic backdoor detection across different network architectures is challenging}.

% Hence, we need to identify more potential Trojan clues from both $\mathbf{w}_{l}$ and $b_t$ to enable robust backdoor detection.

% , and the distribution of $\mathbf{w}_{l}$ and $b_t$ varies from different model structures

\subsection{Data-free Parameter Multidimensional Anomaly Indicators}

To uncover Trojan clues for robust backdoor detection, we design a set of indicators to extract abnormal multidimensional features from the final-layer parameters $\mathcal{C}_{\theta^c}$. From existing studies \cite{fields2021trojan,fu2023freeeagle} and our analysis in Sec. \ref{sec:background}, we can find that there are both possible Trojan clues in the weights $\mathbf{W}\in\mathbb{R}^{D\times K}$ and biases $\mathbf{B}\in\mathbb{R}^{K}$ of $\theta^c$. However, they may vary on model architectures, dataset distributions, and backdoor attacks.
Hence, instead of relying on a single heuristic, we quantify the final layer's parameters ($\mathbf{W}$ and $\mathbf{B}$) from diverse yet complementary dimensions with 13 major indicators to comprehensively identify potential backdoor anomalies:
% Hence, we mainly design 12 major parameter anomaly indicators calculated on $\mathbf{W}$ and/or $\mathbf{B}$ to comprehensively identify potential anomalies from multiple perspectives:

First, we introduce 4 statistical characteristic indicators to describe the fundamental distribution and central tendency of the weight vectors for each class:
\begin{enumerate}[leftmargin=*, noitemsep]
\item \textbf{Weight Mean}: $\mathbf{I}^{\text{WM}}=\{\mu_i\}_{i=1}^K=\{\frac{1}{D}\sum^D_{j=1}W_{i,j}\}_{i=1}^K$.
It reveals the direction of systematic bias: a backdoor class may exhibit a significantly higher or lower mean value, indicating a global shift in its weight distribution compared to benign classes;%系统性偏向性，
\item \textbf{Absolute Weight Mean}: $\mathbf{I}^{\text{AWM}}=\{|\mu_i|\}_{i=1}^K$. It quantifies the overall magnitude of this directional bias, regardless of its sign, helping to identify classes with abnormally strong global weight activations;%整体偏向性，量化偏向强度
\item \textbf{Variance of Weight}:  $\mathbf{I}^{\text{VW}}=\{\frac{1}{D}\sum^D_{j=1}(W_{i,j}-\mu_i)^2\}_{i=1}^K$. It reflects the concentrated nature of the weight distribution. A very high variance might indicate unstable feature dependencies due to conflicts between trigger features and original image features;%特征重要性分布的集中性
\item \textbf{Standard Variance of Weight}: $\mathbf{I}^{\text{SVW}}=\{\sigma_i\}_{i=1}^K$.
It provides a more intuitive measure of dispersion in the same units as the weights, directly indicating the typical fluctuation amplitude of weight values;%权重波动的直观幅度，离散度
\end{enumerate}

Second, we design 3 geometric characteristic indicators that relates to the model's confidence and the geometric positioning of the class in the feature space:
\begin{enumerate}[leftmargin=*, noitemsep, start=5]
\item  \textbf{L1 Norm of Weight}: $\mathbf{I}^{\text{L1}}=\{\mu_i(|W_{i,j}|)\}_{i=1}^K$. It reflects the total absolute contribution strength of all weights for a class, where an abnormal value may indicate an overly strong or weak response;
\item \textbf{L2 Norm of Weight}: $\mathbf{I}^{\text{L2}}=\{\|W_i\|_2\}_{i=1}^K$. It measures the geometric distance of the class's weight vector from the origin, which is intrinsically linked to the distance of its decision boundary. Anomalous norms can signal classes with abnormally strong or weak overall influence;%该指标能够衡量整体权重强度与分类边界的几何距离
\item \textbf{Weight Energy}: $\mathbf{I}^{\text{WE}}=\{\text{Softmax}(\frac{1}{D}\sum^D_{j=1}W^2_{i,j})\}_{i=1}^K$. It reflects the relative importance of classes. A backdoor class might possess disproportionately high weight energy, signifying that the model allocates excessive representational capacity to trigger-related features.
% \item \textbf{Decision distance}: $\mathbf{I}^{\text{DS}}=\{\sum^K_{j=1,j\neq i}\frac{|b_j-b_i|}{\|W_j-W_i\|_2}\}_{i=1}^K$. It quantifies decision boundary differences between classes. Backdoor attacks can distort local geometry, making the boundary around the target class abnormally far from others, which this indicator aims to quantify.
\end{enumerate}

Third, we involve 3 indicators that focus on the bias term and its combination with weights, directly probing the model's prior inclination towards certain classes:
\begin{enumerate}[leftmargin=*, noitemsep, start=8]
  \item \textbf{Bias}: $\mathbf{I}^{\text{B}}=\{b_i\}_i^K$. It represents an explicit prior logit added to a class. An abnormally high bias for a specific class is a straightforward yet strong signature of a backdoor, as the attacker can directly manipulate this term to favor the target class;
  \item \textbf{Summarization of Weight and Bias}: $\mathbf{I}^{\text{SWB}}=\{b_i+\sum_j^DW_{i,j}\}_i^K$.
  It reflects the baseline classification score that provides a holistic measure of the class's inherent activation level, which may be skewed by a backdoor;
  \item \textbf{Summarization of Average Weight and Bias}: $\mathbf{I}^{\text{AWB}}=\{\mu_i^{\text{norm}}+b_i^{\text{norm}}\}_i^K$, where $\text{norm}$ is the Min-Max normalization operation. It helps analyze the source of classifier imbalance by showing whether a class's high baseline score stems predominantly from a large bias or from a consistently high average weight contribution.
\end{enumerate}

Finally, we design 2 advanced indicators to capture structural or semantic anomalies in the decision space:

\begin{enumerate}[leftmargin=*, noitemsep, start=11]

    \item  \textbf{Weight Similarity}: $\mathbf{I}^{\text{WS}}=\{\frac{1}{K}\sum^{K}_{j=1}(1-\frac{W_i^\top W_j}{\|W_i\|_2 \|W_j\|_2})\}_i^K$. It the average cosine distance from one class's weight vector to all others. A backdoor class's weight vector, being tuned to a specific trigger pattern, may become an outlier, showing abnormal directional similarity to other classes;
    \item \textbf{Weight Certainty}: DNN models typically have aleatoric (caused by the inherent uncertainty of training data) and epistemic (caused by the uncertainty of model knowledge) classification uncertainty \cite{sensoy2018evidential}. By injecting poison data, a backdoor forces the model to form an overly certain, simplistic association from trigger to target, thereby reducing the epistemic uncertainty for the target class. Hence, following \cite{sensoy2018evidential}, we design the indicator of weight certainty by reversing the epistemic uncertainty as follows:
    \begin{equation}
      \mathbf{I}^{\text{WC}}=\{1- \text{norm}(I_i^{\text{EU}})\}_{i=1}^K=\{1- \text{norm}(\frac{K}{K+\sum_{j=1}^D W_{i,j}})\}_{i=1}^K,
    \end{equation}
    where $I_i^{\text{EU}}$ is the epistemic uncertainty of the class $i$.  A higher weight certainty (lower epistemic uncertainty) for a class may be a telltale sign of a backdoor implant.
\end{enumerate}

Furthermore, given a major indicator $\mathbf{I}$, we further extend it with 4 metrics for uncovering more potential clues:

\begin{enumerate}[leftmargin=*, noitemsep]
    \item \textbf{Z-Score}:
    \begin{equation}
        \mathbf{I}^{\text{ZS}}=\{\frac{I_i-\mu(\mathbf{I})}{\sigma(\mathbf{I})}\}_i^K.
    \end{equation}
    It is also called standard score and indicates the extent to which the mean value of features in a class deviates from the global mean value;
    \item \textbf{Normalized Absolute Difference}:
    \begin{equation}
        \mathbf{I}^{\text{NAD}}=\{\frac{|I_i-\mu(\mathbf{I})|}{\max(\mathbf{I})-\min(\mathbf{I})}\}_i^K.
    \end{equation}
    It measures the proportion of the deviation degree of each class's indicator mean to the full distance, which can reflect asymmetries in the distribution of indicators;
    \item \textbf{Interquartile Range (IQR) Bounds}: $\{\mathbf{I}^{\text{IQU}},\mathbf{I}^{\text{IQL}}\}$. We use the IQR to measure $\mathbf{I}$ and obtain an upper bound $\mathbf{I}^{\text{IQU}}$ and lower bound $\mathbf{I}^{\text{IQL}}$, respectively,  as follows:
    \begin{align}
        \text{IQR}&=Q_3(\mathbf{I})-Q_1(\mathbf{I}),\\
        \mathbf{I}^{\text{IQU}} &= \mathbf{I}-Q_3 - 1.5*\text{IQR}.\\
        \mathbf{I}^{\text{IQL}} &= Q_1 - 1.5*\text{IQR}-\mathbf{I}.
    \end{align}
\end{enumerate}

With the above 4 extended metrics, we input 12 major indicators into them and totally obtain 60 indicators. Besides, we further summarize the normalized z-score of $\mathbf{I}^{\text{WM}}$ and $\mathbf{I}^{\text{B}}$ as an additional indicator $\mathbf{I}^{\text{WBZ}}=\text{norm}(\mathbf{I}^{\text{ZS}}(\mathbf{I}^{\text{WM}}))+\text{norm}(\mathbf{I}^{\text{ZS}}(\mathbf{I}^{\text{B}}))$ and extend it using $\mathbf{I}^{\text{NAD}}$, which combines the kurtosis of weights and biases to capture anomalies, similar to $\mathbf{I}^{\text{SWB}}$. Finally, we conduct 62 parameter indicators. This ensemble of indicators forms a rich signal set for detecting parameter-space artifacts left by backdoor implants.

% A clean model typically exhibits relative uniformity across these indicators for all classes. In contrast, a backdoor model is expected to manifest significant statistical deviations in the target class across multiple—if not all—of these indicator dimensions.

We further mine combinations of different indicators as a robust Trojan clue matrix $\mathcal{I}^N=\{\mathbf{I}_i\}$ to configure DFBScanner. Since these indicators may be redundant, backdoor-specific, or even inhibit backdoor detection, using a mined $\mathcal{I}^N$ can eliminate the anomalous biases introduced by different backdoor triggers and attack strategies. Hence, we analyze their saliency for detection to selectively integrate indicators as the matrix $\mathcal{I}^N$ using 4 unsupervised/supervised methods on a set of backdoor models, including isolation forest (IForest), mutual information (MutualInfo), logistic regression with L1 regularization (L1-LR), and recursive feature elimination (REF). We employ those methods to identify $\mathcal{I}^N$ for each dataset or network architecture and evaluate their detection accuracy to determine $\mathcal{I}^N$. During the detection stage, DFBScanner only needs to extract values of the configured $\mathcal{I}^N$ from $\theta^c$ to perform backdoor detection.

% \item \textbf{Decision Distance}: $\mathbf{I}^{\text{DD}}=\{\sum_{i=1,i\neq j}^{K}\frac{b_i-b_j}{\|W_i-W_j\|_2}\}^K_i$. This indicator reflects the average distance between each class's decision boundary relative to other classes.

 % (we discuss how we design these indicators and their validity in Appendix Sec. \ref{sec:iva}):

% Specifically, given a desired length $N$ of $\mathcal{I}^N$, we analyze all indicators' importance score to backdoor detection and select top-$N$ indicators to construct $\mathcal{I}^N$. To calculate this importance score, we employ 4 different methods, including isolation forest (IForest), mutual information (MutulInfo), logistic regression with L1 regularization (L1-LR), and recursive feature elimination (REF). IForest is an unsupervised method and can capture nonlinear indicator relationships through a tree structure. MutulInfo provides this importance score by calculating the mutual information value of each indicator and binary label. This method can directly reflect the statistical correlation between indicators and labels, but it ignores the importance of combining indicators. L1-LR forces unimportant indicator coefficients to zero through L1 regularization. It is mainly suitable for addressing linear separability problems. In REF, we employ IForest as the base model to capture nonlinear relationships among indicators. This method iteratively eliminates the least important indicator until the remaining number of indicators is less than $N$.

%Note that since the indicator selection is one-time, we do not need to be care

\subsection{Lightweight Backdoor Detection}
\label{sec:detect}

Given the configured Trojan clue matrix $\mathcal{I}^N$ for a model $\mathcal{M_\theta}$, DFBScanner calculates the detection score $s^\mathcal{M}$ of all classes as follows:
\begin{equation}
    s^\mathcal{M}=\frac{1}{N}\sum_i^N\mathcal{I}^N(\theta^c),
\end{equation}
where $\mathcal{I}^N(\theta^c)\in[0,1]^{K\times N}$. With $s^\mathcal{M}$, DFBScanner performs backdoor detection with two stages: 1) answer whether the input model is poisoned; and 2) answer which class is poisoned when the model is poisoned. In the first stage, a straightforward approach is to use a threshold to distinguish benign and backdoor scores, similar to \cite{fields2021trojan, fu2023freeeagle, wang2024mm}. However, different triggers and attack methods may lead to inconsistent scales in detection scores (as discussed in Sec. \ref{sub:performance}). This results in using a single threshold to distinguish normal and abnormal scores has a high FPR across different attacks. Furthermore, the differences in architectures and the intrinsic data imbalance within datasets can also introduce different degrees of systematic bias in the detection score. Hence, we consider the similarity of parameter distributions between clean and backdoor models since backdoor implantation leads to a certain difference in the final-layer parameter distribution between backdoor and clean models. In the configuration stage, we use a set of clean models to obtain an average clean score distribution $\overline{s}^C$. During detection, DFBScanner calculates the cosine similarity of $s^\mathcal{M}$ and $\overline{s}^\mathcal{C}$ and analyzes whether the model $\mathcal{M}$ is poisoned as follows:
\begin{equation}\label{equ:cos}
\text{sim}(s^\mathcal{M},\overline{s}^C)=\frac{\sum_i^Ks_i^\mathcal{M}\overline{s}_i^\mathcal{C}}{\sum_i^K(s_i^\mathcal{M})^2\sum_i^K(\overline{s}_i^\mathcal{C})^2}<\lambda,
\end{equation}
where $\lambda$ is a similarity threshold to distinguish the clean and backdoor models, which we set according to different model architectures and datasets with a set of backdoor and clean models. Once the above equation is satisfied, in the second stage, DFBScanner identifies the class that has the largest score as the poison class:
\begin{equation}\label{equ:detection}
    i^{\text{backdoor}}=\mathop{\text{argmax}}\limits_{0\leq i<K} \, s_i^M.
\end{equation}
DFBScanner operates exclusively via final-layer parameter analysis, obviating all model inference and input processing to achieve portable, low-cost deployment.

\section{Evalution}
To evaluate the effectiveness and efficiency of DFBScanner, we design the following research questions:

\begin{itemize}[leftmargin=*, noitemsep]
  \item \textbf{RQ1}: How efficient and accurate is DFBScanner in detecting backdoor attacks compared with other SOTA approaches?
    \item \textbf{RQ2}: How effective is DFBScanner with different settings of datasets and model architectures?
    \item \textbf{RQ3}: How can we bypass DFBScanner?
\end{itemize}

\subsection{Experiment Setup}

\subsubsection{Backdoor Detection Benchmark}

We construct a large-scale backdoor detection benchmark to evaluate DFBScanner:

\textbf{Datasets}: We focus on 4 image classification datasets, which cover different scales of labels (10, 10, 43, and 200):

\begin{itemize}[leftmargin=*, noitemsep]
    \item \textbf{MNIST} \cite{lecun1998gradient} is a widely used dataset for handwritten digit recognition. It consists of 60,000 training and 10,000 testing grayscale images (28×28) of digits ranging from 0 to 9;
    \item \textbf{CIFAR-10} \cite{krizhevsky2009learning} is a dataset built for image classification tasks. It contains 60,000 small, labeled images (32×32) across 10 distinct classes;
    \item \textbf{GTSRB} \cite{houben2013detection} is a dataset built for traffic sign classification tasks. It contains 43 categories of traffic signals and is commonly divided into 39,209 training samples and 12,630 test samples (32×32);
    \item \textbf{TinyImageNet} \cite{le2015tiny} (TINet) is a colored image dataset for object recognition. It is downscaled from ImageNet \cite{deng2009imagenet} and contains 200 classes, each of which has 500 training images, 50 validation images, and 50 test images (64×64).
\end{itemize}

\textbf{Model Architectures}: Considering the potential impact of different architectures on backdoor detection, we selected 12 different networks with different depths and architectures:
\begin{itemize}[leftmargin=*, noitemsep]
    \item MNIST: \textbf{CNN2} (which has 2 ConvBlocks (each consisting of a 3$\times$3 convolution, a BatchNorm (BN), and a ReLU layer) and 2 FC layers \cite{gu2017badnets}) and \textbf{LeNet5} (LT5)\cite{lecun1998gradient};
    \item CIFAR-10 and GTSRB: \textbf{CNN6} (3 ConvBlock$\times$2+maxpool and 2 FC layers), \textbf{Resnet18} (R18) \cite{he2016deep}, \textbf{PreactResnet-18} (PR18) \cite{he2016deep}, \textbf{GoogleNet} (GNet) \cite{szegedy2017inception}, \textbf{VGG16-bn} (V16) \cite{simonyan2014very}, and \textbf{Efficientnet-B3} (ENet)\cite{tan2019efficientnet};
    \item TinyImageNet: \textbf{PreactResnet18}, \textbf{Mobilenet-v3-large} (MV3)\cite{howard2019searching}, \textbf{VGG19-bn} (V19), and \textbf{ViT-16-b} (ViT) \cite{dosovitskiy2020image}.
\end{itemize}
% \noindent We describe details of these models in Appendix \ref{sub:model}.

\textbf{Backdoor Attacks}: We study backdoor detection both in the offline and online backdoor injection scenarios. In the online scenario, we study two open-sourced bit-flip backdoor attacks: \textbf{TBT} \cite{rakin2020tbt} and \textbf{HPT} \cite{bai2022hardly}. In the offline scenario,  we employ the following 10 different backdoor attacks both in all-to-one and all-to-all forms:
\begin{itemize}[leftmargin=*, noitemsep]
    \item \textbf{4 input-space attacks}: Badnet \cite{gu2017badnets}, TrojanNN \cite{liu2018trojannn}, Blended \cite{chen2017blended}, and Blind \cite{bagdasaryan2021blind}. Note that different from other attacks, Blind is injected by training pipeline manipulation;
    \item \textbf{6 feature-space attacks}: Inputaware (IA) \cite{nguyen2020inputaware}, Wanet \cite{nguyen2020wanet}, ISSBA \cite{li2021ssba}, Lira \cite{doan2021lira}, Low frequency (LF) \cite{zeng2021rethinking}, and BppAttack (Bpp)\cite{wang2022bppattack}.
\end{itemize}

Class imbalances inherent in the datasets can also lead to uneven decision spaces, which may interfere with the  performance of DFBScanner. Hence, we first implant the above backdoor attacks on all classes of 4 datasets as much as possible in the all-to-one form to conduct a comprehensive benchmark. In this benchmark, we train 11, 21, 25, and 21 clean models per architecture on MNIST, CIFAR10, GTSRB, and TinyImageNet. For each attack per architecture, we implant it on all classes of MNIST, CIFAR10, and GTSRB and on the class set of $[0,5,\cdots,i*5,\cdots,195]$ of TinyImageNet. While models trained on CIFAR10 and GTSRB are injected with all 10 attacks, models trained on MNIST are only injected with BadNet and on TinyImageNet are only injected with 7 attacks, including Badnets, TrojanNN, Blended, BPP, LF, IA, and Wanet. In total, we construct 382 clean models and 4,320 all-to-one backdoor models as the \textbf{full benchmark}.

\begin{figure}[!t]
    \centering
    \includegraphics[width=0.75\linewidth]{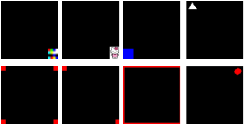}
    \caption{Different trigger patches.}
    \label{fig:triggers}
    \vspace{-5mm}
\end{figure}

Besides, we further train 120 all-to-all backdoor models on CIFAR10 and GTSRB using 6 architectures and 10 attacks. For each attack, we follow a loop permutation to generate target and source class pairs, \textit{i.e.}, each class $k\in K$ is a poison target class with a source class $(k+1)\ \text{mod}\ K$. Furthermore, we introduce 8 different trigger patches in Fig. \ref{fig:triggers} and use Badnet to poison four (3, 5, 7, 9) and five (8, 16, 24, 32, 40) classes of ENet, CNN6, GNet, R18, and P18 on CIFAR10 and GTSRB, respectively and generate 360 backdoor models. We also consider adaptive patches \cite{qi2023adappatch} to train 44 backdoor models on CIFAR10 (all classes) and GTSRB (12 different classes) using R20 and V16, which can achieve latent-separable attacks.
In the online scenario, following the default setting of TBT and HPT, we train 40 backdoor models on CIFAR10 using R18 and V16 with a quantization level of 8-bit. Finally, we introduce 2 adaptive attacks against DFBScanner to train 240 backdoor models. In summary, we construct totally 5,124 backdoor models and 382 corresponding clean models to evaluate DFBScanner.

To enable rapid analysis, we extract a subset of backdoor models as the \textbf{small benchmark}. Specifically, we included all models of MNIST; For CIFAR10, GTSRB, and TinyImageNet, we randomly selected 11, 13, and 11 clean models per architecture and 3, 12, 10 backdoor models with a random-selected target class per attack-architecture pair, respectively. There are 210 clean models and 1200 backdoor models in this benchmark.

\subsubsection{Baselines}

Our experiments employ 7 advanced post-training detection methods as baselines: NC \cite{wang2019neural}, FeatureRE \cite{wang2022rethinking}, DF-TND \cite{wang2020practical}, DQ \cite{fields2021trojan}, Freeeagle \cite{fu2023freeeagle}, MM-BD \cite{wang2024mm}, and BARBIE \cite{zhang2025barbie}. While NC and FeatureRE require data access, other methods are fully data-free; DQ and DFBScanner utilize parameter analysis, whereas others are implemented based on gradient optimization; among these methods, BARBIE can only answer whether the model is backdoored, but cannot identify the backdoor target class. Since these gradient-based methods require estimating backdoor anomalies or trigger patterns in all classes, they are typically highly time-consuming. Hence, following the setting in MM-DB, we reduced the complexity of NC, FeatureRE, and DF-TND by only estimating each putative target class on TinyImageNet and GTSRB. This optimization yields a negligible impact on their detection accuracy, but can accelerate evaluation efficiency.

\subsubsection{Implementation Details}

\begin{table}[t!]
    \centering
    \footnotesize
    \caption{Model training configurations in our experiments}
    \label{tab:trainingconfig}
    \begin{tabular}{c|cccc}
    \hline
        & MNIST & CIFAR10 & GTSRB & TinyImageNet \\\hline
         optimizer & SGD & SGD & SGD & SGD \\
         batch size & 128 & 128 & 128 & 64 \\
         epochs & 100 & 100 & 50 & 100 \\
         learning rate & 1e-3 & 1e-3 & 1e-3 & 1e-3\\
         poison rate & 0.1 & 0.1 & 0.1 & 0.1 \\\hline
    \end{tabular}
    \vspace{-6mm}
\end{table}

We mainly employ BackdoorBench \cite{wu2022backdoorbench} with default configurations in Table \ref{tab:trainingconfig} to construct this benchmark. The ASR of most backdoor models in our benchmark is over than 90\%. Following \cite{fu2023freeeagle}, we further trained 160 clean models and 864 backdoored models (with 0 and last label as backdoor targets per dataset) as a \textit{configuration set} to determine these detection configurations in DFBScanner and other baselines, e.g., the matrix $\mathcal{I}^N$ and the threshold $\lambda$ in DFBScanner, benign boundaries in BARBIE. Note that although NC, DQ, and MM-BD all employ a fixed confidence threshold to detect backdoor across a set of fixed architecture-dataset pairs, we found that once these pairs are changed, their thresholds are no longer optimal. Hence, to achieve a fair comparison, we only re-optimized their thresholds per architecture-dataset pair to maximize their detection performance.

Besides directly configuring DFBScanner with all 62 indicators $\mathcal{I}_{A}^N$ ($N=62$), we further employ IForest, MutualInfo, L1-LR, and REF provided in the PyOD tool \cite{zhao2019pyod} to select indicators to construct optimal matrices, which can obtain the highest F1 according to Equ. (\ref{equ:detection}) with the minimum number $N$ of indicators at the overall- ($\mathcal{I}_{U}^N$), dataset- ($\mathcal{I}_{D}^N$), and model-level ($\mathcal{I}_{M}^N$), respectively. $\mathcal{I}_{U}^N$ is the best matrix for all models in this configuration set, $\mathcal{I}_{D}^N$ is optimal for these models trained on the same dataset, and $\mathcal{I}_{M}^N$ is optimal for these models with the same architecture on the same dataset. Additionally, we use these clean models to obtain $\overline{s}^\mathcal{C}$ for each $\mathcal{I}^N$ and use a Gaussian process-based minimization algorithm\footnote{\url{http://scikit-optimize.github.io}} with the objective of maximizing the F1 score of detecting models to identify the best detection threshold $\lambda$ for each $\mathcal{I}^N$. We configure our DFBScanner with these four matrices $\mathcal{I}_{A}^N$, $\mathcal{I}_{U}^N$, $\mathcal{I}_{D}^N$, and $\mathcal{I}_{M}^N$ as DFBScanner-A, DFBScanner-U, DFBScanner-D, and DFBScanner-M, respectively.

\subsubsection{Evaluation Metrics}
We employ the attack success rate (ASR) and benign accuracy (BA) to evaluate the performance of backdoor models. Similar to FreeEagle \cite{fu2023freeeagle}, we evaluate the detection accuracy of DFBScanner by TPR and FPR. All clean and backdoor models are trained and detected on an Intel i9-12900K CPU with/without 2 NVIDIA RTX3090 GPUs.

\begin{table*}[t!]
\centering
\footnotesize
\caption{Performance comparisons on different datasets (backdoor/benign models). The time is the average detection time per model. For a fair comparison, the time cost does not include the time for loading datasets and model parameters. TPR/FPR are in percentages (\%).}
\label{tab:per}
\resizebox{\textwidth}{!}{
\begin{threeparttable}
\begin{tabular}{c|cc|cc|cc|cc|cc}
\toprule
\multirow{2}{*}{Method} & \multicolumn{2}{c|}{MNIST (20/22)} & \multicolumn{2}{c|}{CIFAR10 (180/66)} & \multicolumn{2}{c|}{GTSRB (720/78)} & \multicolumn{2}{c|}{TinyImagenet (280/44)} & \multicolumn{2}{c}{Total (1200/210) } \\\cline{2-11}
                       & TPR/FPR  & Time   & TPR/FPR     & Time     & TPR/FPR  & Time    & TPR/FPR   & Time    & TPR/FPR     & Time   \\\hline
NC\cite{wang2019neural}\tnote{$\dagger$}&\textbf{100/0}&230.7s&52.78/18.18&383.2s&54.16/12.82&4,382.8s\tnote{$*$}&69.64/15.91&$>$24h\tnote{$*$}& 45.33/13.81&-\\
FeatureRE\cite{wang2022rethinking}\tnote{$\dagger$}&75/0&281.6s&26.67/13.64&380.6s& 37.08/14.1&2,907.5s\tnote{$*$}&69.29/15.91 &50,835s\tnote{$*$}&43.67/12.86 & 13,402s\\\hline
DF-TND\cite{wang2020practical}\tnote{$\dagger$}&40/50&24.9s&20.0/4.54&93.3s&20.83/8.97& 1,058.5s\tnote{$*$}&26.79/2.3& 40,706s\tnote{$*$}&22.42/10.48&9,970s\\
DQ \cite{fields2021trojan}& 15/45.5&\textbf{0.14ms}& 56.67/7.58&\textbf{0.18ms}&   75.69/8.97&\textbf{0.2ms}& 42.5/4.55&\textbf{0.37ms}&   64.08/11.43&\textbf{0.23ms}\\
FreeEagle\cite{fu2023freeeagle}\tnote{$\dagger$}&35/13.63 & 6.36s&26.67/18.18 & 74.6s&46.12/10.26 & 311.2s&24.29/6.82&1,778.1s & 37.92/12.38            & 597.9s            \\
MM-DB\cite{wang2024mm}\tnote{$\dagger$} & 60.0/22.7& 2.0s&33.33/28.79 & 16.3s&41.11/17.95 & 66.4s &34.29/15.91&1,810.1s &38.67/21.43&456.4s \\
BARBIE\cite{zhang2025barbie}\tnote{$\dagger\ddagger$} & 95.00/15.00 & 91.0s  &86.19/62.33  & 176.7s &86.67/61.54  &3,626.1s   &92.14/68.18  &19,235s & 87.85/59.05 &6107.7s\\
\hline
DFBScanner-A & 100/4.55 & 10.63ms  &87.78/4.55 &10.96ms &98.47/3.85 & 129.9ms &94.29/4.55 &2839ms &95.92/4.23 & 728.1ms \\ % None
DFBScanner-U & 100/4.55 &10.27ms&88.33/3.03            &10.48ms&98.47/3.85            &129.1ms&95.36/0         &2805ms& 96.25/2.86           &719.8ms\\
DFBScanner-D & \textbf{100/0} &2.64ms& 88.33/3.03 &10.43ms &99.03/2.56 & 3.59ms&96.43/\textbf{0}&560.4ms& 96.83/1.90&132.7ms\\%None

DFBScanner-M & \textbf{100/0} & \underline{1.17ms} &\textbf{88.33/1.52} &\underline{1.14ms} &\textbf{99.03/0}& \underline{0.995ms}&\textbf{97.86}/2.27&\underline{1.13ms}&\textbf{97.17/0.95}&\underline{1.06ms}\\%None #422590047121048
\hline\hline
&\multicolumn{2}{c|}{MNIST (20/22)}&\multicolumn{2}{c|}{CIFAR10 (600/126)}&\multicolumn{2}{c|}{GTSRB(2580/150)}& \multicolumn{2}{c|}{TinyImagenet (1120/84)}& \multicolumn{2}{c}{Total(4320/382)}\\\hline
DFBScanner-M&100/0&\underline{1.17ms}&\textbf{90.16/7.14}&\underline{1.14ms}&\textbf{98.41/0.67}&\underline{0.995ms} &\textbf{98.21/7.14} &\underline{1.13ms}&\textbf{97.22/4.19}&\underline{1.05ms}\\\bottomrule
\end{tabular}
\begin{tablenotes}
  \item[$*$] denotes the time cost for all classes, but we only estimate the putative target class;
  \item[$\dagger$] denotes running on GPU, others only rely on CPU;
  \item[$\ddagger$] denotes that it can only detect whether the model is backdoored, but cannot identify the backdoor target class.
\end{tablenotes}
\end{threeparttable}
}
\vspace{-3mm}
\end{table*}

\subsection{Experimental Results}

\subsubsection{Backdoor Detection Performance}
\label{sub:performance}
To answer RQ1 and RQ2, we evalauted DFBScanner and other baselines on the small benchmark for saving times, and then further evaluated DFBScanner on the full backdoor model set. Table \ref{tab:per} shows the detection accuracy and time cost of our DFBScanner compared with existing methods. Since NC is designed primarily for patch-based backdoor attacks, it shows a high performance for BadNet, but fails to detect other complex backdoor attacks. NC and FeatureRE require the training set for trigger reversion, resulting in substantial time overhead. DQ detects backdoors only according to average weights $\mathbf{I}^{\text{WM}}$, thereby achieving the fastest detection speed. However, as discussed in Sec. \ref{sec:key}, not all backdoor attacks will amplify $\mathbf{w}_l$ well. Hence, in the backdoor-agonistic scenario, the performance of DQ is not stable. For example, DQ can achieve 100/9.09\% and 100/0\% TPR/FPR for GNet on CIFAR10 and GTSRB, respectively, but fails to detect attacks against PR18, ViT on TinyImageNet, and models on MNIST. DF-TND, FreeEagle, MM-DB, and BARBIE all use random noise as input and perform gradient optimization algorithms to reverse triggers or backdoor anomalies.
MM-BD is designed to detect arbitrary backdoor patterns. However, MM-BD was only evaluated on a few model architectures (e.g., ResNet, VGG, and MobileNet) and does not analyze more architectures, resulting in limited generalization of its conclusions. From our evaluation results, we can see that MM-BD outperforms DF-TND, but shows comparable efficacy to FreeEagle. BARBIE achieves a high TPR, but its FPR is also extremely high. The main reason may be that our benchmark contains diverse clean models trained from scratch with different seeds, leading BARBIE's decision boundary to be less robust \cite{somepalli2022can}.

Compared to existing methods, DFBScanner obtains significantly excellent backdoor detection accuracy and time cost. DFBScanner-U is configured with 55 indicators. Compared with DFBScanner-A, DFBScanner-U demonstrates slight advantages in accuracy and time cost, but it still maintains relatively high time costs. DFBScanner-D enhances both detection accuracy and efficiency, since the number of configured indicators $N$ is less than 62.
On TinyImageNet, DFBScanner-D is configured with 22 indicators and obtains a 2.14\% TPR improvement with a 0\% FPR over DFBScanner-A. DFBScanner-M deepens into the model level to search for the best $\mathcal{I}^N$. It further improves the detection accuracy while significantly reducing detection time. On MNIST, CIFAR10, and GTSRB, DFBScanner-M maintains equivalent TPR while further reducing FPR and detection time. On TinyImageNet, its detection time is reduced from 560.4 ms to 1.13 ms. Note that DQ and DFBScanner are executed only on CPU, whereas other methods rely on GPUs. We also deploy DFBScanner in Raspberry Pi 5b where DFBScanner obtains an average detection time of 4.84 ms for backdoor detection. In Table \ref{tab:datafree}, we further evaluate the efficiency of DFBScanner against DFBA, a very fast backdoor attack \cite{cao2024data}. While DFBA can inject the backdoor into a model within around 10 ms, DFBScanner can detect this attack only within around 0.23 ms, thereby effectively preventing online injection of this attack. Overall, DFBScanner achieves the best TPR and FPR with sufficiently low latency for practical run-time detection.

\renewcommand{\arraystretch}{1.0}
\begin{table}[t!]
\centering
\footnotesize
\caption{TPR/FPR against the DFBA attack \cite{cao2024data}. The attack time is DFBA's average time of injecting a backdoor into a model.}
\label{tab:datafree}
\resizebox{\columnwidth}{!}{
\begin{tabular}{c|c|cccc|c|c}
\toprule
\multirow{2}{*}{Method}&\multirow{2}{*}{} & \multicolumn{2}{c}{ResNet18} & \multicolumn{2}{c|}{VGG16} & \multirow{2}{*}{\makecell{Attack \\ time}} & \multirow{2}{*}{\makecell{Avg. Detection\\time}} \\\cline{3-6}
 &&TPR$\uparrow$ & FPR$\downarrow$&TPR$\uparrow$ &FPR$\downarrow$& &\\\hline
DQ    & \multirow{3}{*}{C}&\textbf{10/10}  &\textbf{0/10}  &1/10 &\textbf{0/10} &\multirow{3}{*}{\makecell{10.7$\pm$ \\0.5ms}} &\textbf{0.13$\pm$0.02ms} \\
 MM-BD &&0/10&2/10   &1/20&4/10  & &28.92$\pm$1.9s \\
 DFBScanner && \textbf{10/10}&\textbf{0/10} & \textbf{10/10}&\textbf{0/10} &  &\underline{0.22$\pm$0.2ms}\\\hline
DQ    & \multirow{3}{*}{G}&\textbf{22/22} &\textbf{0/22}  &0/22 &2/22  &\multirow{3}{*}{\makecell{10.9$\pm$\\0.7ms}} &\textbf{0.13$\pm$0.02ms} \\
MM-BD  & & 0/22&1/16   &1/22 &3/16 & & 59.1$\pm$3.2s \\
 DFBScanner  && \textbf{22/22}&\textbf{0/22} & \textbf{22/22}&\textbf{0/22} &  &\underline{0.25$\pm$0.2ms}\\\hline
\end{tabular}
}
\vspace{-2mm}
\end{table}

\begin{table}[t!]
\centering
\footnotesize
\caption{TPR (\%) of DFBScanner across different architectures. M- denotes MNIST.}
\label{tab:attack}
\resizebox{\columnwidth}{!}{
\begin{tabular}{c|cccccc}
\hline
Dataset& R18& PR18 & ENb3 & V16 & GNet & CNN6 \\\hline
CIFAR10 & 100 & 99.0 & 96.0 & 80.0 & 100 & 66.0 \\
GTSRB & 100 & 99.5 & 98.6 & 96.5 & 97.2 & 98.6 \\\hline\hline
  & MV3 & PR18 & V19 & \multicolumn{1}{c||}{ViT} & M-LT5 & M-CNN2\\\hline
Tiny & 97.9 & 100 & 98.2 & \multicolumn{1}{c||}{96.8} & 100 & 100 \\\hline
\end{tabular}}
\vspace{-2mm}
\end{table}

Table \ref{tab:per} also illustrates the detection performance of DFBScanner-M on the full benchmark. DFBScanner-M maintains consistent performance across different classes. This demonstrates that DFBScanner's performance is not affected by the class imbalance problem. Additionally, Table \ref{tab:attack} details the TPR performance achieved by DFBScanner-M across different network architectures. On CIFAR10, among different architectures, DFBScanner-M has a relatively low performance on V16 and CNN6 due to anomaly dilution across multiple FC layers. While CIFAR10's limited classification space exacerbates this dilution, a larger classification space (e.g., on GTSRB and TinyImageNet) can alleviate this dilution.

Through this large-scale evaluation, we demonstrate that DFBScanner obtains a superior backdoor detection performance in terms of accuracy, speed, and resource overhead. %without any input data processing
%From the attack aspect, DFBScanner-M fails to detect Blind, Bpp, and LF attacks with CNN6 and Bpp attacks with VGG16-bn.
\subsubsection{Detecting Attacks with Various Trigger Patches}
We further consider backdoor attacks with various trigger patches that have different patch sizes, locations, color perturbations, and also adaptive generation \cite{qi2023adappatch}. We aim to analyze the generalization of DFBScanner against diversified responses between different model architectures and trigger patches.
Table \ref{tab:badnet} demonstrates the detection performance of MM-BD, DQ, and DFBScanner. DFBScanner is configured with $\mathcal{I}_M^N$, $\overline{s}^\mathcal{C}$, and $\lambda$ of DFBScanner-M. MM-DB's backdoor detection process depends on the ability of the attack to make the target class logit abnormally high. However, due to those patch differences, diverse patches lead to various marginal distribution changes. So, using a fixed threshold for each architecture to perform detection results in a lower TPR of MM-BD. MM-BD only achieves a clearly better TPR than others on the CNN6 models on CIFAR10. DFBScanner shows a more robust and accurate performance across different patches contributed by its used robust Trojan clues.

\begin{table}[t!]
\centering
\footnotesize
\caption{TPR of DFBScanner against Badnet attacks with different patch patterns. C and G denote CIFAR10 and GTSRB, respectively. The last two columns are TPR of detecting adapt-patch attacks \cite{qi2023adappatch}.}
\label{tab:badnet}
\resizebox{\columnwidth}{!}{
\begin{tabular}{c|c|ccccc||cc}
\hline
Method & & ENet & CNN6 & GNet & R18 & PR18 & R20 & V16 \\\hline
MM-DB & \multirow{3}{*}{C} &28/32 &\textbf{14}/32&2/32&8/32&15/32 & 2/10 & 0/10\\
DQ &  &19/32 &1/32&\textbf{32}/32&31/32&0/32 &\textbf{10/10}& \textbf{10/10}\\
DFBScanner & &\textbf{32/32}  &8/32&\textbf{32/32} &\textbf{32/32}&\textbf{31/32} &\textbf{10/10} &\textbf{10/10}\\%None
\hline
MM-DB & \multirow{3}{*}{G} &17/40 &34/40 &7/40 &31/40 &\textbf{38}/40 &3/12 &0/12  \\
DQ &  &31/0  &34/40&\textbf{40/40}&\textbf{40/40}&0/40 &11/12 &\textbf{12/12}\\
DFBScanner &  &\textbf{40/40}  &\textbf{40/40} &\textbf{40/40} & \textbf{40/40}&37/40&\textbf{12/12}&\textbf{12/12} \\\hline
\end{tabular}
}
\vspace{-2mm}
\end{table}

\begin{table}[t!]
\footnotesize
\centering
\caption{Detection performance against bit-flip attacks.}
\label{tab:bitflip}
\begin{tabular}{c|cccc|c}
\hline
\multirow{2}{*}{Method} & \multicolumn{2}{c}{Resnet18} & \multicolumn{2}{c|}{VGG16} & \multirow{2}{*}{Avg. Time} \\\cline{2-5}
 &TPR$\uparrow$ & FPR$\downarrow$&TPR$\uparrow$ &FPR$\downarrow$&\\\hline
 MM-DB &7/20&1/10  &10/20&1/10 & 30.32s \\
 DQ    &13/20&\textbf{0/10} &11/20&\textbf{0/10} & \textbf{0.12ms} \\
 DFBScanner & \textbf{20/20}&\textbf{0/10} & \textbf{20/20}&\textbf{0/10} &  \underline{0.22ms}\\\hline
\end{tabular}
\vspace{-2mm}
\end{table}

\subsubsection{Backdoor Detection of Bit-Flip Attacks}
We also study the online attack scenario with bit-flip attacks. Table \ref{tab:bitflip} shows the detection results of MM-DB, DQ, and DFBScanner. Since flipping the find-layer parameters is typically the optimal approach to implant backdoor attacks with the minimal number of flipped bits, most of the existing bit-flip attacks follow this approach. Through indicator analysis, we find that the indicator $\mathbf{I}^{\text{L2}}$ or $\mathbf{I}^{\text{WE}}$ has sufficient information to distinguish benign and backdoor models and identify the poison class, but $\mathbf{I}^{\text{WM}}$ can not. Hence, DFBScanner can achieve a 100\% TPR and 0\% FPR, and its average detection time per model is 0.22 ms, which is sufficient to ensure real-time attack detection.

\subsubsection{Backdoor Detection of All-to-All Attacks} We use the configurations ($\mathcal{I}^N$, $\lambda$, and $\overline{s}^C$) of DFBScanner-M to detect all-to-all models in the first stage. Table \ref{tab:all2all} shows the detection results. In all‐to‐all backdoor attacks, instead of mapping to a single target class, poisoned samples coming from the same source class have a corresponding target class. This attacks lead to the final-layer parameters of each class all contain anomalies. So DFBScanner accurately distinguishs benign and backdoor models by measuring their cosine similarity.

\begin{table}[t!]
\footnotesize
\centering
\caption{TPR of DFBScanner for detecting all-to-all attacks.}
\label{tab:all2all}
\begin{tabular}{c|cccccc}
\hline
Dataset & ENet & CNN6 & GNet & R18 & PR18 &VGG \\\hline
CIFAR10 &10/10  &10/10 &10/10 &10/10& 10/10&10/10 \\%\hline
GTSRB  &10/10  &10/10  &10/10 & 10/10&10/10&10/10 \\\hline
\end{tabular}
\vspace{-5mm}
\end{table}

\subsubsection{Backdoor Detection in Practical Scenario}

A practical detection scenarios is that the defender downloads a set of models from online that may contain some backdoored models, and she need to scan these models to identify potential backdoored models. Therefore, we evaluate DFBScanner and other baselines under this scenario in which the model set is contaminated by a few of backdoored models. We randomly select 1, 2, and 3 backdoored models per architecture-dataset pair to combine with clean models. Since we have no the configuration set to determine the detection $\lambda$ and $\overline{s}^\mathcal{C}$, we calculate cosine similarities among all models' detection score $s$ and use the z-score method to filter out outliers in their average similarities. Table \ref{tab:practical} demonstrates DFBScanner's detection performance. We can see that in practical scenarios, DFBScanner still achieves a high TPR and low FPR.

\begin{table}
  \caption{Detection performance of DFBScanner in practical scenarios where DFBScanner has no reference configuration set to determine the threshold $\lambda$ and average clean score $\overline{s}^\mathcal{C}$.}
  \label{tab:practical}
  \centering
  \begin{tabular}{cc|cc|cc|cc}
    \hline
    \multicolumn{2}{c|}{MNIST}  & \multicolumn{2}{c|}{CIFAR10} & \multicolumn{2}{c|}{GTSRB} & \multicolumn{2}{c}{TinyImageNet}\\\hline
    TPR & FPR &TPR & FPR &TPR & FPR &TPR & FPR \\\hline
    2/2&0/22 & 6/6 &1/126 &6/6 &0/150 &4/4&0/84\\
    4/4&0/22 &12/12 &1/126 &11/12 & 0/150&8/8&0/84\\
    6/6&0/22 &17/18 &0/126 & 17/18&0/150 &11/12 &0/84\\
    \hline
  \end{tabular}
  \vspace{-4mm}
\end{table}

\subsection{Ablation Study}

\subsubsection{Indicator Validity Analysis}
\label{sec:iva}
\begin{figure*}
    \centering
    \includegraphics[width=0.95\textwidth]{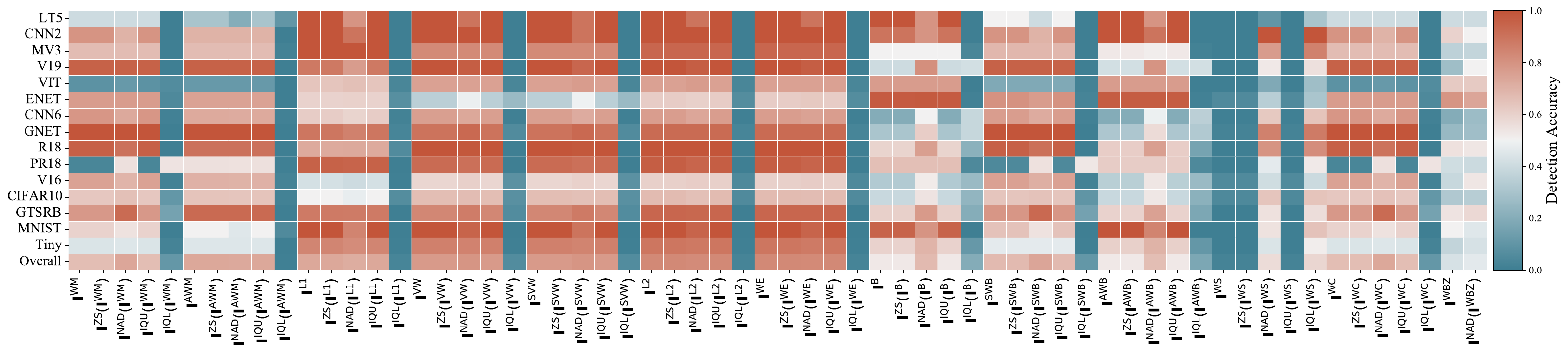}
    \vspace*{-2mm}
    \caption{Backdoor detection accuracy of parameter anomaly indicators using maximum anomaly indices.}
    \label{fig:heatmap}
    \vspace*{-4mm}
\end{figure*}

We evaluate the backdoor detection accuracy of all proposed parameter indicators on 780 backdoor models across all attacks, models, and datasets, determined by Equ. \eqref{equ:detection}. Fig. \ref{fig:heatmap} shows that most indicators are valid (ACC$>$0.5), although their effectiveness varies by architecture and dataset. CIFAR10 yields lower ACC (max 67.3\%) compared to GTSRB (94.3\%) and TinyImageNet (89.3\%), due to its smaller label space constraining final-layer capacity. In contrast, TinyImageNet’s 200 classes facilitate easier backdoor implantation via local fine-tuning. GTSRB’s structured traffic signs further reduce implantation difficulty. Across architectures, $\mathbf{I}^{\text{L2}}$ and $\mathbf{I}^{\text{WE}}$ perform best in most models.

\subsubsection{Indicator Selection}
\label{sec:iselect}
\begin{figure}
    \centering
  \begin{minipage}[c]{1\columnwidth}
    \centering
    \includegraphics[width=0.7\columnwidth]{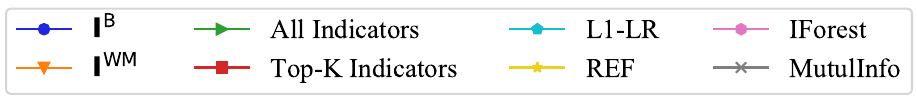}
  \end{minipage}
  \vspace{-4mm}

  \begin{minipage}[c]{0.9\columnwidth}
    \centering
    \subfloat[MNIST]{\includegraphics[width=0.45\columnwidth]{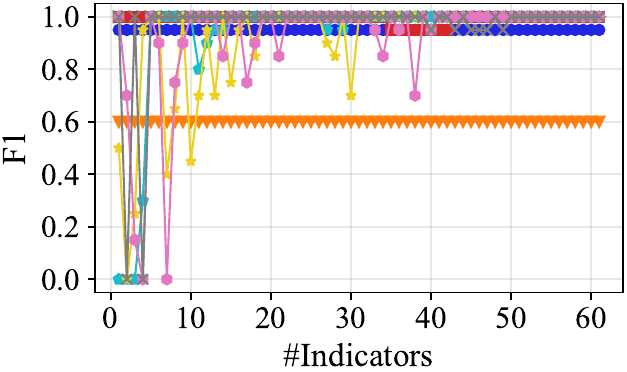}}
    \hfill
    \subfloat[CIFAR10]{\includegraphics[width=0.45\columnwidth]{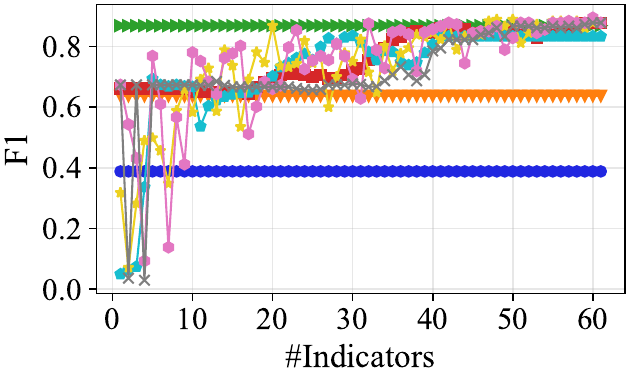}}
  \end{minipage}
    \vspace{-4mm}

  \begin{minipage}[c]{0.9\columnwidth}
    \centering
    \subfloat[GTSRB]{\includegraphics[width=0.45\columnwidth]{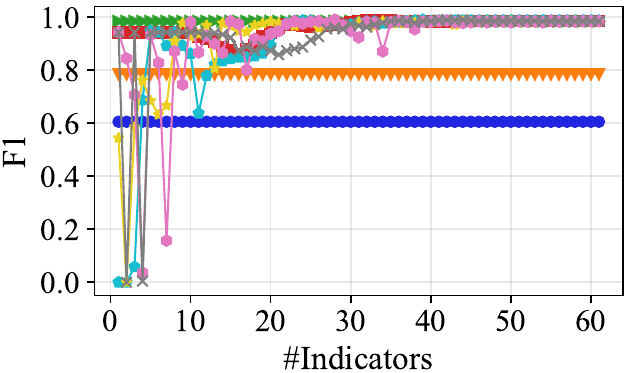}}
    \hfill
    \subfloat[TinyImageNet]{\includegraphics[width=0.45\columnwidth]{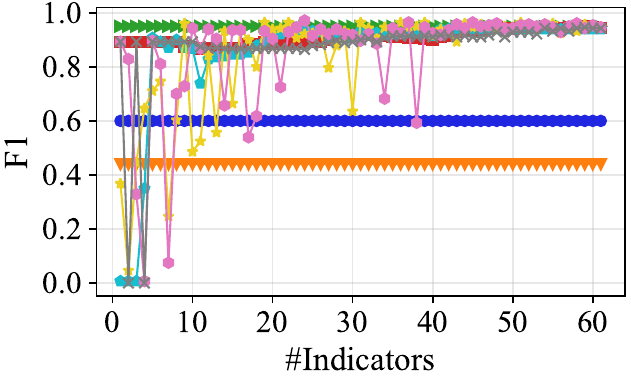}}
  \end{minipage}

\caption{Backdoor detection F1-score curve with different numbers $N$ of indicators generated by indicator selection.}
\label{fig:featureselection}
   \vspace{-4mm}
\end{figure}

We use 2 single indicators ($\mathbf{I}^{\text{WM}}$ and $\mathbf{I}^{\text{B}}$), and all 62 indicators (`All Indicators')  to construct 3 baseline clue matrices. Besides, we also sort these indicators according to their accuracy in Fig. \ref{fig:heatmap} and select them to construct detection matrices $\mathcal{I}^N$ with different numbers of indicators (Top-K Indicators). We compare these four baselines with 4 matrix selection methods (IForest, MutulInfo, L1-LR, and REF) at the overall, dataset, and model levels. Fig. \ref{fig:featureselection} illustrates the F1-score of $\mathcal{I}^N$ with different $N$ at the dataset level. $\mathbf{I}^{\text{WM}}$ and $\mathbf{I}^{\text{B}}$ can only yield limited detection performance. `All Indicators' can enable sufficiently ideal accuracy and stability for backdoor detection. This corroborates the essential effectiveness of our proposed method. The F1 of `Top-K Indicators' improves with the increase in $N$, and can even exceed the performance of `All Indicators'. While these 4 matrix selection methods find more compact $\mathbf{I}$ with a higher F1 score, compared to other baselines. Furthermore, performing indicator selection at the model level can produce a performance better than at the dataset level, as shown in Table \ref{tab:per}. In summary, combining multiple indicators can yield robust Trojan clues, but how to select and combine is impacted both by dataset characteristics and network architectures.

\subsubsection{Model Cosine Similarity}
\label{sec:msim}
\begin{figure*}
\centering
    \subfloat[CIFAR10]{\includegraphics[width=0.85\textwidth]{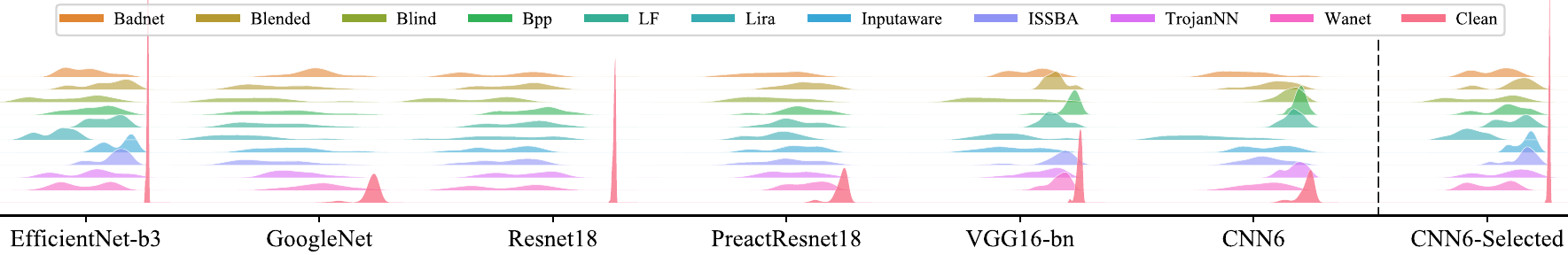}}
    \vspace{-3mm}
    \subfloat[GTSRB]{\includegraphics[width=0.85\textwidth]{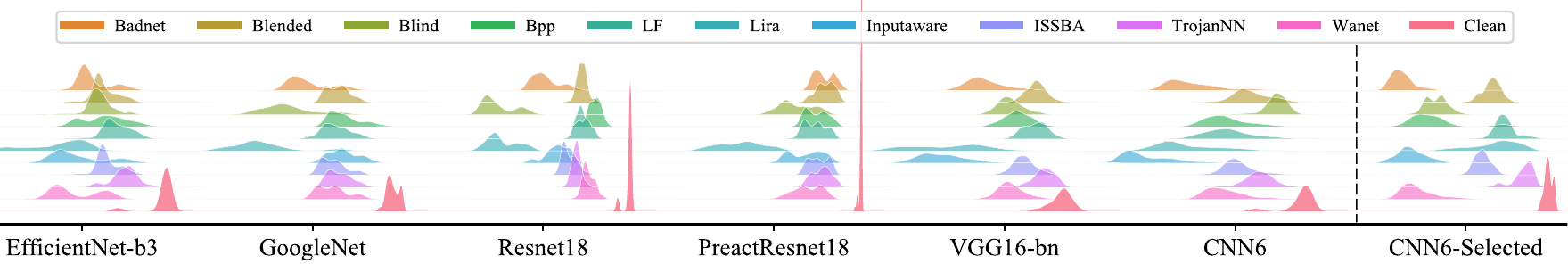}}
    \caption{Cosine similarity of the anomaly score between benign and backdoor models using all indicators and selected indicators. CNN6-Selected denotes the anomaly score using the selected clue matrix; other models all use all indicators.}
    \label{fig:sim}
    \vspace*{-2mm}
\end{figure*}

We analyze the distinguishability between clean and backdoor models using cosine similarity, following Equ. \eqref{equ:cos}. We extract all indicators from models trained on CIFAR10 and GTSRB and obtain their anomaly score. We then calculate the cosine similarity between clean models and between clean and backdoor models, and show their probability density function (PDF) in Fig. \ref{fig:sim}. Overall, most backdoor models are distinguishable from their clean models, especially in high-capacity architectures such as ENet, GNet, R18, PR18, and V16. This separability arises because backdoor perturbations $\delta$ alter classification weights $\mathbf{W}$, disrupting their original alignment with latent features $\mathbf{f}(x)$. While clean models show compact anomaly score distributions, backdoor models exhibit greater variability due to diverse triggers and strategies. CNN6 on CIFAR10 is less distinguishable, as its compact classification capability requires broader feature-space adjustments rather than local tuning. By computing cosine similarity of anomaly scores from a selected clue matrix, it enhances separability by removing redundant or negative indicators. The high mutual similarity among clean models enables a reliable threshold $\lambda$ that effectively discriminates backdoor models under both all-to-one and -all attacks, achieving under 5\% FPR in Table \ref{tab:per}.

\subsubsection{Impact of Configuration Set}
We further assess the impact of configuration sets. In the original setup, we trained 2 backdoor models with a fixed traget label per attack-architecture-dataset triple. We further trained 3 additional configuration sets: (1) 1 backdoor model with 1 random target per triple; (2) 3 backdoor models with random targets per triple; and (3) 3 backdoor models with random targets, but trained using a randomly selected half of attacks per architecture-dataset pair. Using these sets, we determine DFBScanner's hyperparameters at the model level. Table \ref{tab:iconfig} demonstrates DFBScanner's TPR and FPR under these configurations. We can see that more backdoor attacks and targets can improve the accuracy of DFBScanner. But, with less backdoor attacks in the configuration set, DFBScanner still obtains a high TPR and low FPR. This demonstrates the robustness of DFBScanner.

\begin{table}[t!]
  \caption{Impact of the configuration set to DFBScanner's performance. `R-' denotes random-selected. The first row uses the original configuration set. All results are are repeated 3 times.}
  \label{tab:iconfig}
  \centering
  \resizebox{\columnwidth}{!}
  {
  \begin{tabular}{cccccc}
    \hline
    \#Attacks & \#Targets &MNIST &CIFAR10 & GTSRB & TinyImageNet\\\hline
    All & 2 & 100.0/0.00 & 90.16/7.14 &98.41/0.67 &98.21/7.14\\\hline
    All & R-1 &100.0/4.55 &89.83/2.38 &99.15/1.33 &97.86/4.76\\\hline
    All & R-3 &100.0/0.00 &91.17/4.76  &99.46/2.00  & 98.57/3.57\\\hline
    R-half & R-3 &100.0/4.55 &88.17/3.17 &98.45/1.33 & 96.79/4.76 \\\hline
  \end{tabular}
  }
  \vspace{-2mm}
\end{table}

\subsection{Possible Adaptive Attack}

To answer RQ3, we investigate two possible adaptive attack that may bypass DFBScanner: (1) the adaptive attacker adds a regularization on parameters $\theta^c$ to the model training loss, similar to \cite{fields2021trojan}; (2) the adaptive attacker employs the classifier part of benign models and freeze its parameters during backdoor injection.

\begin{table}[!t]
\footnotesize
  \caption{DFBScanner's detection performance against adaptive attacks. Note that all metrics are in \%.}
    \label{tab:adaptive}
    \centering
    \resizebox{\columnwidth}{!}{
    \begin{tabular}{c|ccc|ccc}
    \hline
         \multirow{2}{*}{Dataset}&\multicolumn{3}{c|}{Regularization} &\multicolumn{3}{c}{Parameter Frozen}\\\cline{2-7}
         & $\Delta^{\text{BA}}$ & $\Delta^{\text{ASR}}$ & TPR/FPR & $\Delta^{\text{BA}}$ & $\Delta^{\text{ASR}}$ & TPR/FPR \\\hline
        CIFAR10 &-1.44 &-2.74 &65.0/6.06 &-2.73&-9.38&11.7/6.06\\\hline
         GTSRB  & -1.24&-2.19 &56.7/0 &-2.25&-7.92&2.5/0\\\hline
    \end{tabular}}
    \vspace{-5mm}
\end{table}

\textbf{1) Adaptive Attack via Regularizing Parameters}: Similar to \cite{fields2021trojan}, we modify the model training loss function with regularization both on weight and bias as follows:
\begin{equation}
\mathcal{L}=\mathcal{L}_{\text{ce}}+\beta\sum_{i=1}^K\|\mathbf{w}_i-\overline{\mathbf{w}}\|^2+\gamma\sum_{i=1}^K(b_i-\overline{b})^2,
\end{equation}
where $\overline{\mathbf{w}}$ and $\overline{b}$ are the average weights and biases of all classes, $\beta$ and $\gamma$ control the suppression intensity. This loss function forces $\mathbf{w}_l$ and $b_l$ of the backdoor class $l$ to stay close to other classes.

\textbf{2) Adaptive Attack via Freezing Parameters}: In this attack, the attacker first trains a clean model and then freezes its $\theta^c$ to retrain the model using the poison dataset. This method can defend against many backdoor detection methods (e.g., FreeEagle, DQ, and our DFBScanner), but it may have an impact on the classification performance of benign samples since it can only be achieved by optimizing the feature extraction layer.

We train 240 backdoor models on CIFAR10 and GTSRB per adaptive attack and configure DFBScanner with $\mathcal{I}_M^N$. Table \ref{tab:adaptive} illustrates these models' performance and DFBScanner's detection performance. Both adaptive attacks can bypass DFBScanner's detection to some extent. But the second attack can result in a higher decrease of ACC and ASR than the first attack, and DFBScanner can still detect part of the backdoor models. Freezing the last layer is an effective way to evade DFBScanner, but at the cost of higher decreased BA and ASR.

\subsection{Discussion}

We evaluate DFBScanner with a large-scale set of backdoor models on different datasets, network architectures, and backdoor attacks, and demonstrate its superior detection performance in terms of accuracy, efficiency, and robustness. However, it still has several limitations. First, once an adversary with full knowledge of DFBScanner, she can easily evade it by freezing the find-layer parameters. This adaptive attack also introduces performance degradation of models. In this condition, we need to move parameter inspection to upper layers to find potential Trojan clues. Second, we only focus on how to detect backdoor attacks, but how to mitigate found backdoor attacks has not been investigated yet. As studied in \cite{wang2022rethinking, zhang2024exploring}, neuron activation values representing the Trojan behavior in the latent space are orthogonal to others. A potential backdoor mitigation method is that since we have no knowledge of backdoor triggers, we may be able to utilize the abnormal indicators in the last layer to reverse the activation direction of $\|\delta\|$ (see Sec. \ref{sec:key}) via singular value decomposition on $\mathbf{W}$, similar to \cite{phan2024clean}. Considering the orthogonality \cite{zhang2024exploring}, we can perform suppression for any activations in the latent space on this trigger direction. In addition, we can also use this direction to project the exception weights and biases, and correct the last layer of parameters through parameter fine-tuning. We leave this extension as future work.

\section{Conclusion}
In this work, we identify robust Trojan clues in the final-layer parameters of models and propose DFBScanner, a data-free backdoor detection framework based on parameter analysis, to address the robust detection problem across diverse datasets, network architectures, backdoor triggers, and attack strategies. DFBScanner uses 62 anomaly indicators to inspect the final-layer parameters and a lightweight detection method. It accurately detects backdoor models without requiring model inference or input data, achieving over 97\% detection accuracy with approximately 1.0 ms overhead per model in our large-scale backdoor benchmark, enabling practical and efficient deployment.

% Pay more attention on other DNN backdoor attacks against like DRL policies \cite{yu2023spatiotemporal,yu2024spatiotemporal}, NLP translator \cite{},

\bibliographystyle{IEEEtran}
\bibliography{IEEEabrv, ref}

% Generated by IEEEtran.bst, version: 1.14 (2015/08/26)
\begin{thebibliography}{10}
\providecommand{\url}[1]{#1}
\csname url@samestyle\endcsname
\providecommand{\newblock}{\relax}
\providecommand{\bibinfo}[2]{#2}
\providecommand{\BIBentrySTDinterwordspacing}{\spaceskip=0pt\relax}
\providecommand{\BIBentryALTinterwordstretchfactor}{4}
\providecommand{\BIBentryALTinterwordspacing}{\spaceskip=\fontdimen2\font plus
\BIBentryALTinterwordstretchfactor\fontdimen3\font minus
  \fontdimen4\font\relax}
\providecommand{\BIBforeignlanguage}[2]{{%
\expandafter\ifx\csname l@#1\endcsname\relax
\typeout{** WARNING: IEEEtran.bst: No hyphenation pattern has been}%
\typeout{** loaded for the language `#1'. Using the pattern for}%
\typeout{** the default language instead.}%
\else
\language=\csname l@#1\endcsname
\fi
#2}}
\providecommand{\BIBdecl}{\relax}
\BIBdecl

\bibitem{zhang2024badmerging}
J.~Zhang, J.~Chi, Z.~Li, K.~Cai, Y.~Zhang, and Y.~Tian, ``Badmerging: Backdoor
  attacks against model merging,'' in \emph{ACM CCS}, 2024, pp. 4450--4464.

\bibitem{gu2017badnets}
T.~Gu, B.~Dolan-Gavitt, and S.~Garg, ``Badnets: Identifying vulnerabilities in
  the machine learning model supply chain,'' \emph{arXiv preprint
  arXiv:1708.06733}, 2017.

\bibitem{tran2018spectral}
B.~Tran, J.~Li, and A.~Madry, ``Spectral signatures in backdoor attacks,''
  \emph{NeurIPS}, vol.~31, 2018.

\bibitem{chen2019detecting}
B.~Chen, W.~Carvalho, N.~Baracaldo, H.~Ludwig, B.~Edwards, T.~Lee, I.~Molloy,
  and B.~Srivastava, ``Detecting backdoor attacks on deep neural networks by
  activation clustering,'' in \emph{AAAI Workshop}, 2019.

\bibitem{chan2019poison}
A.~Chan and Y.-S. Ong, ``Poison as a cure: Detecting \& neutralizing
  variable-sized backdoor attacks in deep neural networks,'' \emph{arXiv
  preprint arXiv:1911.08040}, 2019.

\bibitem{xu2021detecting}
X.~Xu, Q.~Wang, H.~Li, N.~Borisov, C.~A. Gunter, and B.~Li, ``Detecting ai
  trojans using meta neural analysis,'' in \emph{IEEE S\&P}, 2021, pp.
  103--120.

\bibitem{chen2019deepinspect}
H.~Chen, C.~Fu, J.~Zhao, and F.~Koushanfar, ``Deepinspect: A black-box trojan
  detection and mitigation framework for deep neural networks.'' in
  \emph{IJCAI}, vol.~2, no.~5, 2019, p.~8.

\bibitem{wang2019neural}
B.~Wang, Y.~Yao, S.~Shan, H.~Li, B.~Viswanath, H.~Zheng, and B.~Y. Zhao,
  ``Neural cleanse: Identifying and mitigating backdoor attacks in neural
  networks,'' in \emph{IEEE S\&P}, 2019, pp. 707--723.

\bibitem{qiao2019defending}
X.~Qiao, Y.~Yang, and H.~Li, ``Defending neural backdoors via generative
  distribution modeling,'' in \emph{NeurIPS}, vol.~32, 2019.

\bibitem{wang2022rethinking}
Z.~Wang, K.~Mei, H.~Ding, J.~Zhai, and S.~Ma, ``Rethinking the
  reverse-engineering of trojan triggers,'' vol.~35, pp. 9738--9753, 2022.

\bibitem{ma2024need}
Z.~Ma, Y.~Yang, Y.~Liu, T.~Yang, X.~Liu, T.~Li, and Z.~Qin, ``Need for speed:
  Taming backdoor attacks with speed and precision,'' in \emph{IEEE S\&P},
  2024, pp. 1217--1235.

\bibitem{wang2020practical}
R.~Wang, G.~Zhang, S.~Liu, P.-Y. Chen, J.~Xiong, and M.~Wang, ``Practical
  detection of trojan neural networks: Data-limited and data-free cases,'' in
  \emph{ECCV}, 2020, pp. 222--238.

\bibitem{fu2023freeeagle}
C.~Fu, X.~Zhang, S.~Ji, T.~Wang, P.~Lin, Y.~Feng, and J.~Yin, ``Freeeagle:
  Detecting complex neural trojans in data-free cases,'' in \emph{USENIX
  Security}, 2023, pp. 6399--6416.

\bibitem{zhou2024data}
Q.~Zhou, W.~Luo, Z.~Ye, and Y.~Tang, ``Data-free backdoor model inspection:
  Masking and reverse engineering loops for feature counting,'' in
  \emph{IJCNN}.\hskip 1em plus 0.5em minus 0.4em\relax IEEE, 2024, pp. 1--9.

\bibitem{wang2024mm}
H.~Wang, Z.~Xiang, D.~J. Miller, and G.~Kesidis, ``Mm-bd: Post-training
  detection of backdoor attacks with arbitrary backdoor pattern types using a
  maximum margin statistic,'' in \emph{IEEE S\&P}, 2024, pp. 1994--2012.

\bibitem{zhang2025barbie}
H.~Zhang, Y.~Bai, Y.~Chen, Z.~Ma, and W.~Xu, ``Barbie: Robust backdoor
  detection based on latent separability,'' in \emph{NDSS}, 2025.

\bibitem{sun2025peftguard}
Z.~Sun, T.~Cong, Y.~Liu, C.~Lin, X.~He, R.~Chen, X.~Han, and X.~Huang,
  ``Peftguard: detecting backdoor attacks against parameter-efficient
  fine-tuning,'' in \emph{IEEE S\&P}, 2025, pp. 1713--1731.

\bibitem{cao2024data}
B.~Cao, J.~Jia, C.~Hu, W.~Guo, Z.~Xiang, J.~Chen, B.~Li, and D.~Song, ``Data
  free backdoor attacks,'' \emph{NeurIPS}, vol.~37, pp. 23\,881--23\,911, 2024.

\bibitem{rakin2020tbt}
A.~S. Rakin, Z.~He, and D.~Fan, ``Tbt: Targeted neural network attack with bit
  trojan,'' in \emph{CPVR}, 2020, pp. 13\,198--13\,207.

\bibitem{costales2020live}
R.~Costales, C.~Mao, R.~Norwitz, B.~Kim, and J.~Yang, ``Live trojan attacks on
  deep neural networks,'' in \emph{CVPR}, 2020, pp. 796--797.

\bibitem{dosovitskiy2020image}
A.~Dosovitskiy, L.~Beyer, A.~Kolesnikov, D.~Weissenborn, X.~Zhai,
  T.~Unterthiner, M.~Dehghani, M.~Minderer, G.~Heigold, S.~Gelly \emph{et~al.},
  ``An image is worth 16x16 words: Transformers for image recognition at
  scale,'' \emph{arXiv preprint arXiv:2010.11929}, 2020.

\bibitem{bagdasaryan2021blind}
E.~Bagdasaryan and V.~Shmatikov, ``Blind backdoors in deep learning models,''
  in \emph{USENIX Security}, 2021, pp. 1505--1521.

\bibitem{bai2022hardly}
J.~Bai, K.~Gao, D.~Gong, S.-T. Xia, Z.~Li, and W.~Liu, ``Hardly perceptible
  trojan attack against neural networks with bit flips,'' in \emph{ECCV}.\hskip
  1em plus 0.5em minus 0.4em\relax Springer, 2022, pp. 104--121.

\bibitem{liu2018trojannn}
Y.~Liu, S.~Ma, Y.~Aafer, W.-C. Lee, J.~Zhai, W.~Wang, and X.~Zhang, ``Trojaning
  attack on neural networks,'' in \emph{NDSS}, 2018.

\bibitem{chen2017blended}
X.~Chen, C.~Liu, B.~Li, K.~Lu, and D.~Song, ``Targeted backdoor attacks on deep
  learning systems using data poisoning,'' \emph{arXiv preprint
  arXiv:1712.05526}, 2017.

\bibitem{doan2021lira}
K.~Doan, Y.~Lao, W.~Zhao, and P.~Li, ``Lira: Learnable, imperceptible and
  robust backdoor attacks,'' in \emph{ICCV}, 2021, pp. 11\,966--11\,976.

\bibitem{li2021ssba}
Y.~Li, Y.~Li, B.~Wu, L.~Li, R.~He, and S.~Lyu, ``Invisible backdoor attack with
  sample-specific triggers,'' in \emph{ICCV}, 2021, pp. 16\,463--16\,472.

\bibitem{zeng2021rethinking}
Y.~Zeng, W.~Park, Z.~M. Mao, and R.~Jia, ``Rethinking the backdoor attacks'
  triggers: A frequency perspective,'' in \emph{ICCV}, 2021, pp.
  16\,473--16\,481.

\bibitem{qi2023adappatch}
X.~Qi, T.~Xie, Y.~Li, S.~Mahloujifar, and P.~Mittal, ``Revisiting the
  assumption of latent separability for backdoor defenses,'' in \emph{ICLR},
  2023.

\bibitem{tao2024distribution}
G.~Tao, Z.~Wang, S.~Feng, G.~Shen, S.~Ma, and X.~Zhang, ``Distribution
  preserving backdoor attack in self-supervised learning,'' in \emph{IEEE
  S\&P}, 2024, pp. 2029--2047.

\bibitem{nguyen2020wanet}
T.~A. Nguyen and A.~T. Tran, ``Wanet-imperceptible warping-based backdoor
  attack,'' in \emph{ICLR}, 2020.

\bibitem{hu2020practical}
X.~Hu, Y.~Zhao, L.~Deng, L.~Liang, P.~Zuo, J.~Ye, Y.~Lin, and Y.~Xie,
  ``Practical attacks on deep neural networks by memory trojaning,'' \emph{IEEE
  TCAD}, vol.~40, no.~6, pp. 1230--1243, 2020.

\bibitem{rakin2019bit}
A.~S. Rakin, Z.~He, and D.~Fan, ``Bit-flip attack: Crushing neural network with
  progressive bit search,'' in \emph{ICCV}, 2019, pp. 1211--1220.

\bibitem{nguyen2020inputaware}
T.~A. Nguyen and A.~Tran, ``Input-aware dynamic backdoor attack,''
  \emph{NeurIPS}, vol.~33, pp. 3454--3464, 2020.

\bibitem{wang2022bppattack}
Z.~Wang, J.~Zhai, and S.~Ma, ``Bppattack: Stealthy and efficient trojan attacks
  against deep neural networks via image quantization and contrastive
  adversarial learning,'' in \emph{CVPR}, 2022, pp. 15\,074--15\,084.

\bibitem{zhang2022poison}
J.~Zhang, C.~Dongdong, Q.~Huang, J.~Liao, W.~Zhang, H.~Feng, G.~Hua, and N.~Yu,
  ``Poison ink: Robust and invisible backdoor attack,'' \emph{IEEE TIP},
  vol.~31, pp. 5691--5705, 2022.

\bibitem{huang2022backdoor}
K.~Huang, Y.~Li, B.~Wu, Z.~Qin, and K.~Ren, ``Backdoor defense via decoupling
  the training process,'' \emph{arXiv preprint arXiv:2202.03423}, 2022.

\bibitem{gao2019strip}
Y.~Gao, C.~Xu, D.~Wang, S.~Chen, D.~C. Ranasinghe, and S.~Nepal, ``Strip: A
  defence against trojan attacks on deep neural networks,'' in \emph{ACSAC},
  2019, pp. 113--125.

\bibitem{doan2020februus}
B.~G. Doan, E.~Abbasnejad, and D.~C. Ranasinghe, ``Februus: Input purification
  defense against trojan attacks on deep neural network systems,'' in
  \emph{ACSAC}, 2020, pp. 897--912.

\bibitem{guo2023scale}
J.~Guo, Y.~Li, X.~Chen, H.~Guo, L.~Sun, and C.~Liu, ``Scale-up: An efficient
  black-box input-level backdoor detection via analyzing scaled prediction
  consistency,'' in \emph{ICLR}, 2023.

\bibitem{xiang2020detection}
Z.~Xiang, D.~J. Miller, and G.~Kesidis, ``Detection of backdoors in trained
  classifiers without access to the training set,'' \emph{IEEE TNNLS}, vol.~33,
  no.~3, pp. 1177--1191, 2020.

\bibitem{liu2019abs}
Y.~Liu, W.-C. Lee, G.~Tao, S.~Ma, Y.~Aafer, and X.~Zhang, ``Abs: Scanning
  neural networks for back-doors by artificial brain stimulation,'' in
  \emph{ACM CCS}, 2019, pp. 1265--1282.

\bibitem{tang2021demon}
D.~Tang, X.~Wang, H.~Tang, and K.~Zhang, ``Demon in the variant: Statistical
  analysis of dnns for robust backdoor contamination detection,'' in
  \emph{USENIX Security}, 2021, pp. 1541--1558.

\bibitem{cai2022randomized}
R.~Cai, Z.~Zhang, T.~Chen, X.~Chen, and Z.~Wang, ``Randomized channel
  shuffling: minimal-overhead backdoor attack detection without clean
  datasets,'' in \emph{NeurIPS}, 2022, pp. 33\,876--33\,889.

\bibitem{kolouri2020universal}
S.~Kolouri, A.~Saha, H.~Pirsiavash, and H.~Hoffmann, ``Universal litmus
  patterns: Revealing backdoor attacks in cnns,'' in \emph{CVPR}, 2020, pp.
  301--310.

\bibitem{fields2021trojan}
G.~Fields, M.~Samragh, M.~Javaheripi, F.~Koushanfar, and T.~Javidi, ``Trojan
  signatures in dnn weights,'' in \emph{ICCV}, 2021, pp. 12--20.

\bibitem{yao2020deephammer}
F.~Yao, A.~S. Rakin, and D.~Fan, ``Deephammer: Depleting the intelligence of
  deep neural networks through targeted chain of bit flips,'' in \emph{USENIX
  Security}, 2020, pp. 1463--1480.

\bibitem{huggingface}
H.~Face, ``Hugging face – the ai community building the future,''
  \url{https://huggingface.co}.

\bibitem{chen2021proflip}
H.~Chen, C.~Fu, J.~Zhao, and F.~Koushanfar, ``Proflip: Targeted trojan attack
  with progressive bit flips,'' in \emph{ICCV}, 2021, pp. 7718--7727.

\bibitem{feng2025contrastive}
Y.~Feng, B.~Ma, D.~Liu, Y.~Zhang, W.~Cai, and Y.~Xia, ``Contrastive neuron
  pruning for backdoor defense,'' \emph{IEEE TIP}, vol.~34, pp. 1234--1245,
  2025.

\bibitem{sensoy2018evidential}
M.~Sensoy, L.~Kaplan, and M.~Kandemir, ``Evidential deep learning to quantify
  classification uncertainty,'' \emph{NeurIPS}, vol.~31, 2018.

\bibitem{lecun1998gradient}
Y.~LeCun, L.~Bottou, Y.~Bengio, and P.~Haffner, ``Gradient-based learning
  applied to document recognition,'' \emph{Proceedings of the IEEE}, vol.~86,
  no.~11, pp. 2278--2324, 1998.

\bibitem{krizhevsky2009learning}
A.~Krizhevsky, G.~Hinton \emph{et~al.}, ``Learning multiple layers of features
  from tiny images,'' 2009.

\bibitem{houben2013detection}
S.~Houben, J.~Stallkamp, J.~Salmen, M.~Schlipsing, and C.~Igel, ``Detection of
  traffic signs in real-world images: The german traffic sign detection
  benchmark,'' in \emph{IJCNN}.\hskip 1em plus 0.5em minus 0.4em\relax IEEE,
  2013, pp. 1--8.

\bibitem{le2015tiny}
Y.~Le and X.~Yang, ``Tiny imagenet visual recognition challenge,'' \emph{CS
  231N}, vol.~7, no.~7, p.~3, 2015.

\bibitem{deng2009imagenet}
J.~Deng, W.~Dong, R.~Socher, L.-J. Li, K.~Li, and L.~Fei-Fei, ``Imagenet: A
  large-scale hierarchical image database,'' in \emph{CVPR}, 2009, pp.
  248--255.

\bibitem{he2016deep}
K.~He, X.~Zhang, S.~Ren, and J.~Sun, ``Deep residual learning for image
  recognition,'' in \emph{CVPR}, 2016, pp. 770--778.

\bibitem{szegedy2017inception}
C.~Szegedy, S.~Ioffe, V.~Vanhoucke, and A.~Alemi, ``Inception-v4,
  inception-resnet and the impact of residual connections on learning,'' in
  \emph{AAAI}, vol.~31, no.~1, 2017.

\bibitem{simonyan2014very}
K.~Simonyan and A.~Zisserman, ``Very deep convolutional networks for
  large-scale image recognition,'' \emph{arXiv preprint arXiv:1409.1556}, 2014.

\bibitem{tan2019efficientnet}
M.~Tan and Q.~Le, ``Efficientnet: Rethinking model scaling for convolutional
  neural networks,'' in \emph{ICML}.\hskip 1em plus 0.5em minus 0.4em\relax
  PMLR, 2019, pp. 6105--6114.

\bibitem{howard2019searching}
A.~Howard, M.~Sandler, G.~Chu, L.-C. Chen, B.~Chen, M.~Tan, W.~Wang, Y.~Zhu,
  R.~Pang, V.~Vasudevan \emph{et~al.}, ``Searching for mobilenetv3,'' in
  \emph{ICCV}, 2019, pp. 1314--1324.

\bibitem{wu2022backdoorbench}
B.~Wu, H.~Chen, M.~Zhang, Z.~Zhu, S.~Wei, D.~Yuan, and C.~Shen,
  ``Backdoorbench: A comprehensive benchmark of backdoor learning,''
  \emph{NeurIPS}, vol.~35, pp. 10\,546--10\,559, 2022.

\bibitem{zhao2019pyod}
Y.~Zhao, Z.~Nasrullah, and Z.~Li, ``Pyod: A python toolbox for scalable outlier
  detection,'' \emph{Journal of machine learning research}, vol.~20, no.~96,
  pp. 1--7, 2019.

\bibitem{somepalli2022can}
G.~Somepalli, L.~Fowl, A.~Bansal, P.~Yeh-Chiang, Y.~Dar, R.~Baraniuk,
  M.~Goldblum, and T.~Goldstein, ``Can neural nets learn the same model twice?
  investigating reproducibility and double descent from the decision boundary
  perspective,'' in \emph{CVPR}, 2022, pp. 13\,699--13\,708.

\bibitem{zhang2024exploring}
K.~Zhang, S.~Cheng, G.~Shen, G.~Tao, S.~An, A.~Makur, S.~Ma, and X.~Zhang,
  ``Exploring the orthogonality and linearity of backdoor attacks,'' in
  \emph{IEEE S\&P}, 2024, pp. 2105--2123.

\bibitem{phan2024clean}
H.~Phan, J.~Xiao, Y.~Sui, T.~Zhang, Z.~Tang, C.~Shi, Y.~Wang, Y.~Chen, and
  B.~Yuan, ``Clean \& compact: Efficient data-free backdoor defense with model
  compactness,'' in \emph{ECCV}, 2024.

\end{thebibliography}
\end{document}